\documentclass[format=acmsmall, review=false, screen=true]{acmart}
\usepackage[utf8]{inputenc}
\usepackage{booktabs} % For formal tables

\usepackage[ruled]{algorithm2e} % For algorithms
\usepackage{booktabs}

\usepackage{ragged2e}
\usepackage{courier}
\usepackage{url}
\usepackage{graphicx}
\usepackage{color}
\usepackage{array}
\usepackage{subcaption}
\usepackage[T1]{fontenc}
\usepackage{aecompl}
\usepackage{multirow}

\SetAlFnt{\small}
\SetAlCapFnt{\small}
\SetAlCapNameFnt{\small}
\SetAlCapHSkip{0pt}
\IncMargin{-\parindent}

\newtheorem{resques}{Research Question}

\setcopyright{none}
\begin{document}
% Title portion. Note the short title for running heads
\title[Differentiating High and Low Viability Teams through Team Interaction]{My Team Will Go On: Differentiating High and Low Viability Teams through Team Interaction}
% characterizing? 

% author
%\iffalse
\author{Hancheng Cao}
\email{hanchcao@stanford.edu}
\affiliation{
  \institution{Stanford University}
  \country{United States}}
 
\author{Vivian Yang}
%\email{hanchcao@stanford.edu}
\affiliation{
  \institution{Stanford University}
  \country{United States}}
  
\author{Victor Chen}
%\email{hanchcao@stanford.edu}
\affiliation{
  \institution{Stanford University}
  \country{United States}}
  
\author{Yu Jin Lee}
%\email{hanchcao@stanford.edu}
\affiliation{
  \institution{Stanford University}
  \country{United States}}
 
\author{Lydia Stone}
%\email{hanchcao@stanford.edu}
\affiliation{
  \institution{Stanford University}
  \country{United States}}
 
\author{N'godjigui Junior Diarrassouba}
%\email{hanchcao@stanford.edu}
\affiliation{
  \institution{Texas Tech}
  \country{United States}}
  
\author{Mark E. Whiting}
\email{markew@seas.upenn.edu}
\orcid{0000-0002-6395-7833}
\affiliation{
  \institution{University of Pennsylvania}
  \country{United States}}
  
\author{Michael S. Bernstein}
\orcid{0000-0001-8020-9434}
\email{msb@cs.stanford.edu}
\affiliation{
  \institution{Stanford University}
  \country{United States}}
%\fi

\begin{abstract}
Understanding team viability --- a team's capacity for sustained and future success --- is essential for building effective teams. In this study, we aggregate features drawn from the organizational behavior literature to train a viability classification model over a dataset of $669$ 10-minute text conversations of online teams. We train classifiers to identify teams at the top decile (most viable teams), 50th percentile (above a median split), and bottom decile (least viable teams), then characterize the attributes of teams at each of these viability levels. We find that a lasso regression model achieves an accuracy of $.74$--$.92$ AUC ROC under different thresholds of classifying viability scores.
From these models, we identify the use of exclusive language such as `but' and `except', and the use of second person pronouns, as the most predictive features for detecting the most viable teams, suggesting that active engagement with others' ideas is a crucial signal of a viable team. Only a small fraction of the 10-minute discussion, as little as 70 seconds, is required for predicting the viability of team interaction. This work suggests opportunities for teams to assess, track, and visualize their own viability in real time as they collaborate. 

\end{abstract}





%\thanks{XXX}

\maketitle

%\renewcommand{\shortauthors}{Z. Chen et al.}
\renewcommand{\shortauthors}{Hancheng Cao et al.}

\section{Introduction}

Even high performing teams can be miserable. Whether or not a team succeeds at their goals, their team environment can be joyless, toxic, or even downright malicious. \textit{Team performance} is a widely studied metric of successful collaboration~\cite{mesmer2009information, stewart2006meta, bell2007deep, weidmann2020team}, but it is only one piece of the picture. \textit{Team viability} --- the capacity of a team for sustainable growth and future success~\cite{bell2011} --- is a less studied yet equally critical component of a sustainable, successful team~\cite{mathieu2008team}.

A team with high viability can adapt to internal and external changes~\cite{Cascio2016}. A team which suffers a loss of viability will see its members resist, resent, or even actively undermine their teams~\cite{langfred2007downside, mueller1994societal, sprigg2000production, vallas2003}.
A loss of viability can be so disastrous that the members fracture and do not want to work together again~\cite{whiting2019did}. 
Consequently, it is critical for teams to evaluate and sustain their viability in order to promote continued engagement and future success.
In particular, remote teams are more susceptible to antisocial behavior and developing conflicts~\cite{HindsPamela2003}. As remote collaboration becomes increasingly prevalent and necessary, it becomes even more crucial to study team viability, which reflects both the members' satisfaction with their teammates and the members' behavioral intent to remain in the team~\cite{hackman1980work}.
Since team viability is not necessarily skill-based or task-specific, not just remote teams but any team can strive to improve their viability.

Team viability is typically measured through self-reported questionnaires~\cite{cooperstein2017initial}, limiting its effectiveness to after-the-fact instances of low viability. 
Assessing team viability at early stages of team interaction can offer actionable insights for improving ongoing collaborative efforts, as opposed to collecting retrospective lessons and regrets in the case of post-hoc evaluation. Therefore we ask: can we differentiate more viable and less viable teams based on a glimpse of their interaction? What attributes of team interaction are indicative of high viability? If this were possible, we could aid teams in diagnosing problematic starts and aid them in righting themselves.

To address these questions, we study a dataset of 669 online, 10-minute long, remote team interactions~\cite{whiting2019did, whiting2020parallel}, in which each team completed a collaborative task and a post-hoc team viability instrument~\cite{cooperstein2017initial}. We then use this data to judge the efficacy of particular signals in classifying whether a given interaction has a viability above or below the viability thresholds of top decile, median, or bottom decile. Since we are interested in which features differentiate the behavior of more viable and less viable interactions rather than the particular viability score they achieve, we focus on the task of classification.

We design features drawn from organizational behavior literature and compare multiple machine learning models for classification. For instance, we measure cosine similarity of TF-IDF vectors based on unigrams and bigrams to characterize topic alignment in the conversation, and analyze subjectivity and sentiment of the messages to characterize linguistic output using LIWC (Linguistic Inquiry Word Count)~\cite{somasundaran2010recognizing, tausczik2010psychological}.
Furthermore, features that cannot be easily detected by algorithms, such as psychological safety~\cite{edmondson1999psychological, bradley2012reaping, duggan2017, bradley2012reaping}, were coded by Amazon Mechanical Turk workers. We employ a lasso (L1 regularized) regression model to identify the most predictive features, achieving the highest accuracy of $.92$ AUC ROC for the classification of the top decile viability teams, $.75$ AUC ROC for the task with a median split, and $.74$ AUC ROC for the classification of the lowest decile viability teams. These results remain robust even when controlling for team performance. 

Given these models, we highlight a few behavior patterns that we found as being important signals for viability. First, we find that use of exclusive language, such as `but', `either', `except', and addressing via the second person, such as `you' and `your', are significant predictors of higher viability across all three tasks, which suggest teams whose members actively respond to and engage with the ideas of others are more likely to be viable. Sarcastic humor, anxiety terms, and higher average reading level were additionally significant in distinguishing the top decile teams from the rest, suggesting that the most viable teams are engaging in nuanced and complex interactions, in which context and focus matter. On the other hand, the use of words indicating sadness as defined by LIWC and the negativity text sentiment of the most negative individual were associated with the lowest decile viability teams. 

Many of the significant features in our models, such as those reflecting word choice and sentiment, were computationally derived rather than human-coded. This result hints at the possibility of enabling teams to conduct real-time assessment of their viability. Furthermore, we demonstrate that the relationship between viability and team interaction is consistent over time, yet the pattern differs between the first time a team convenes and when it meets again even if they do not recognize their teammates~\cite{whiting2019did}, which suggests that viability is a dynamic measure that could be monitored continuously. Finally, we find that models using features computationally derived from the first 70 seconds of an interaction can achieve a comparable level of performance to the models utilizing full 10 minute long interaction, implying that signals of team viability arise even in the first few moments of interaction. 

While previous works have focused on predicting and understanding the determinants of team performance~\cite{mesmer2009information, bell2007deep}, team viability is an equally critical goal for promoting effective collaboration. Evaluating viability is especially important in remote teams because they lack a shared context~\cite{olson2000distance}, are more likely to have difficulty developing mutual understanding~\cite{FussellKrauss1992}, are less cohesive than face-to-face groups~\cite{strausandmcgrath}, and exhibit more competitive behavior~\cite{purdy2000}, which render them more prone to developing conflicts~\cite{HindsPamela2003}. In this paper, we present models for assessing viability via classification of online team interaction and explore potential signals of team viability in text interaction data. We make our classification algorithm available for others to use.\footnote{\url{viability.stanford.edu}}

\section{Background and Related work}
Teams are complex dynamic systems that develop as members interact and respond to the unfolding of situational demands over time in a given context~\cite{kozlowski2007team}. On one hand, a positive team experience can help members come together to produce a whole that is greater than the sum of its parts. On the other hand, a negative team experience can sabotage success to the point where members vow to never work together again. Given the critical role that teams and teamwork play in our economy and society, how might we better understand the viability of any given team?

While team viability and team performance are correlated with a strong and positive relationship for teams completing routine activities~\cite{bell2011}, teams that perform well can still have disastrously low viability, as exemplified by the eventual fracture between Kahneman and Tversky~\cite{lewis2017undoing}. Moreover, viability and performance also tend to have a weaker relationship when the team encounters dynamic situations that require different strategies and skills~\cite{kozlowski1998training}. Thus, team viability should be considered and studied as a distinct feature from team performance.

\subsection{Team Viability}
When team viability is absent or lost to such an extent that the team chooses not to work together in the future, a phenomenon called team fracture occurs~\cite{whiting2019did}. Literature shows that interventions for improving the psychological state of a team and its members can be effective~\cite{lacerenza2018team}. If we can recognize that a team has low viability or is dropping in viability, it would be possible to try to help them before they fail~\cite{hjalmarson2012forming}. Thus, there exists an important need for being able to recognize that a team needs help.

Currently, there are two ways that this need may be addressed, and both of them have significant drawbacks. The first mechanism for recognizing that a team needs help is through self-reporting from individual team members or the team as a collective. For example, an individual may raise complaints about their team to the team manager or other parties. This mechanism encounters three problems. First, self-reporting depends on the \textit{ability} of an individual to recognize low or falling team viability. People are often remarkably bad at recognizing their own state~\cite{mesmer2009information}, and may not realize the decline of team viability in the moment. Second, the \textit{responsibility} of speaking up about problems with team viability entirely falls on the reporting individual. Social loafing and diffusion of responsibility are obstacles to self-reporting even if an individual recognizes problems, particularly when a task is unattractive. Third, \textit{potential consequences} present significant disincentives for the reporting party. Self-reporting may potentially lead to interventions and improvement~\cite{lacerenza2018team}, but it could also lead to total disintegration of collaboration or the ostracization of the reporting individual from the team and even the entire organization~\cite{daim2012exploring}. Indeed, even in the best case scenario where the overall viability of a team increases as a result of an individual's report, the social status of that individual is still likely to suffer~\cite{o2008professional}. Because of all of these reasons, individuals often may not act to prevent the impending failure of a team.

The second existing mechanism for recognizing that a team needs help is through external involvement, often taking the form of third-party actors, such as managers, coaches, and counselors, being called to step in via processes such as quarterly team assessments~\cite{maslach2012making}. External involvement allows for the external evaluation of a team, and third-party actors often have incentive to report and act upon problems with a team's viability~\cite{bell2011team}. These factors account for some of the difficulties of relying on self-reporting as a solution. However, external involvement has its own drawbacks. First, introducing an evaluating party who is not directly in the team introduces potential inaccuracies in their evaluation~\cite{mathieu2008team}. When this party is related to the team, e.g. in the case of a team manager~\cite{barrick1998relating}, they may bring biases and misaligned incentives of their own about their team in competition with other teams, or the characteristics of individual members of their team~\cite{bell2011team}. When this evaluating party is entirely separate from the team, e.g. in the case of a counsellor or a team-building professional~\cite{aube2005team}, they may not be able to understand the complete context in which the team lives, and their evaluations may be distorted as well. Second, even if we disregard whether an accurate evaluation of the team is possible, an external human evaluation of the team costs resources, which presents a barrier to entry for smaller teams or organizations. Finally, by the time a third-party actor is summoned to deal with team issues, it may already be too late to matter~\cite{kozlowski1999developing}.

Current ways of recognizing that a team needs help with viability are not ideal --- a gap exists in the literature, with regards to a systematic investigation of the features of team viability and its possible predictability. Team interactions have been the subject of countless rounds of observation and analysis. One point of consensus in prior research is that predicting team behaviour is difficult, often because whether the team dynamics are effective is not obvious even to the team members themselves~\cite{mesmer2009information}. In this work, we set out to answer a related research question:

\begin{resques}[RQ\ref*{resques:canWe}] \label{resques:canWe}
 Can we differentiate high and low viability teams by analyzing team interactions?
\end{resques}

\subsection{Antecedents of team viability}\label{sec:antecedents}
Existing literature explores predicting team fracture and team performance, but not team viability, using field studies and crowdsourcing. Certain features emerge as strong predictors of team performance in field studies, such as team minimum agreeableness and team mean conscientiousness, openness to experience, collectivism, and preference for teamwork~\cite{bell2007deep}. Interesting insights derived from prior research are that linguistic cues and conversational patterns extracted from the first 20 seconds of a team discussion are predictive of whether it will be a productive discussion~\cite{niculae2016conversational}, and that the type of task matters for the consistency of team success or fracture~\cite{whiting2019did}. 
Notably, existing literature has focused on a number of broad feature categories, where there is an overlap between the categories. In this paper, we do not aim to identify patterns on the level of the categories, but we do draw upon trends in these categories to select our features.

\subsubsection{Work Pattern Features}
From proposing ideas to implementing decisions, equal participation is a key aspect of constructive teams~\cite{constructivedisc}. Well-balanced participation is a challenge in virtual, chat-based discussions because they lack in-person cues. For example, in offline interactions, individuals can demonstrate that they are listening to a person speak by nodding their head in agreement, or maintaining eye contact. However, in online interactions, individuals can only convey their listening by sending written responses~\cite{Kriplean2012}. This limitation opens the door to many problematic scenarios as it can be exploited by internet trolls who take pleasure in upsetting others, as well as socially-insensitive individuals who simply neglect to respond~\cite{Kirman2012}. When a person does not see evidence that they are being heard, they feel dissatisfied and frustrated. Such frustration makes people less willing to participate and contribute, sometimes disengaging completely.

A team's performance is also an antecedent to its viability~\cite{mach2010differential, Barrick1998}, but the relationship is complicated by the stresses of achieving performance in some situations~\cite{bell2011team}. Producing high quantities of ideas can be equated to high performance since teams that are diverse, insightful, and creative are better able to cultivate a rich flow of ideas. Another way to evaluate a team's performance is its ability to reach conclusions efficiently, as cohesive teams tend to collectively process ideas and converge to a final decision more quickly than non-cohesive teams~\cite{dionne2010}. 

\subsubsection{Semantic Features}
Constructive communication has an ameliorative effect on groups, reducing conflict and increasing cooperation~\cite{barsade2002}. For example, positive interactions induce a pleasant overall mood, which is associated with greater cognitive effort and more constructive methods of dealing with disagreement. The foundations of a healthy team atmosphere are also commonly established during a team’s first interactions, since initial communication and interaction behaviors set the standard for team culture~\cite{suchan2001}. In particular, friendly introductions and greetings between group members create social bonds, which lay the foundation for benevolence and trust~\cite{greenberg2007}. To sustain this trust, team members often engage in supportive communication in accomplishing a task (e.g. celebrating achievements, expressing appreciation of each other's contributions). Additionally, since online discussions do not convey the same richness of emotion and reaction that face-to-face communication enables, they are susceptible to misunderstandings when utterances contain sarcasm or irony~\cite{muresan2016}. Due to the use of sarcasm, what might look like a negative utterance is, in fact, a positive one (and vice versa). Consequently, sarcastic or ironic utterances online can affect team members’ perceptions of positivity and interaction behaviors. 

Rudeness and aggression can quickly sabotage a healthy conversation (e.g., ad hominem attacks, name-calling, knee-jerk contradiction)~\cite{civilizedDiscussion}. Instead, healthy team dynamics center on agreeableness, respect, and support. From softening the perceived force of a message to allowing all parties to save face, politeness can shape the course of online interactions for the better~\cite{convogoneawry}. Inter-member civility also encourages social connection and rapport, fostering amicable conversation between members and helping groups maintain cohesion throughout a discussion.

\subsubsection{Topic Features}
Some topical contexts naturally lend themselves to antisocial behavior and discord. Specifically, topics like politics, religion, and ideology are breeding grounds for conflict~\cite{convogoneawry}. Religion and philosophy account for over half of the conflict on Wikipedia~\cite{Kittur2009}, while 14\% of Americans say they have been harassed online due to their political views~\cite{duggan2017}. Moreover, people tend to feel more strongly about ideological issues and take matters personally~\cite{catsrule}. Such emotional investment makes individuals more susceptible to friction and hostility, with decision-making becoming more emotional than logical. 

\subsubsection{Word Choice Features}
Psychological safety refers to a shared belief held by teammates that a team is safe for interpersonal risk taking~\cite{edmondson1999psychological}. Task conflicts that occur in a psychologically safe environment have been found to improve creativity and decision making without damaging interpersonal relations or interactions between teammates~\cite{bradley2012reaping}. In this way, a highly productive team may be highly critical but also highly psychologically safe. In the opposite direction, a highly critical team that is not psychologically safe can transform how teammates respond to challenges. A multitude of factors can result in teams not being psychologically safe, for example name-calling, physical threats, dismissive remarks, and social punishments associated with rejecting the group norms~\cite{duggan2017, bradley2012reaping}. In 2017, 41\% of Americans said they have suffered online harassment, 27\% said they have been called offensive names online, and 22\% say that someone intentionally embarrassed them online~\cite{duggan2017}. Victims of such behaviors often experience mental or emotional stress, and are unwilling to work with their perpetrators again. Such stress is particularly common in online mediums, where anonymity emboldens perpetrators to use words that they likely do not use in in-person interactions. These situations motivate another research question: 

\begin{resques}[RQ\ref*{resques:whichFeatures}] \label{resques:whichFeatures}
 Which behaviors are most strongly associated with high vs. low viability teams?
\end{resques}

\subsection{The durability of team viability antecedents}
Team interactions evolve over time~\cite{hackman2011collaborative} --- what indicates a positive relationship early in a team interaction, might indicate a problematic one if the team is expected to be very familiar due to months of close collaboration. Though rare, studies of longitudinal interactions in teams show compounding effects of low viability~\cite{larson2019team}. And thus the question of how repeated episodes of work might expose unexpected qualities of team viability. Intriguingly, the dataset analyzed in this work has the unique property of involving repeated interactions between the same team without them knowing it, as well as interactions when they did know they were working with the same prior collaborators. While the interactions in the data are only 10 minutes long each, their repeated nature may shed light on another research question: 

\begin{resques}[RQ\ref*{resques:howLong}] \label{resques:howLong}
 How does the historic collaboration context influence team viability?
\end{resques}

\subsection{Early warnings}
When studying events with final outcomes, it can be helpful to understand whether that final result can be identified by some patterns or behaviors early in the process rather than after the outcome has already occurred. ``Thin slicing'' studies show how relatively small and relatively early interactions can be powerful predictors of long-term outcomes --- famously by accurately predicting which marriages will fail years later~\cite{gottman2000timing,carrere1999predicting}, or the grades at the end of a semester~\cite{jung2016coupling} --- by making judgements based on only a few minutes of interactions. Understanding viability in the future, as opposed to after a team has already come to have low viability, would turn predicting viability into something strikingly useful for teams in practice. This goal motivates a final research question to study: 
\begin{resques}[RQ\ref*{resques:howShort}] \label{resques:howShort}
    How little duration must be observed to make accurate predictions of team viability?
\end{resques}

In this section we have mapped out the challenges in classifying team viability and motivated a range of theoretically and practically interesting questions to try to answer. We next move in to explaining how we achieve the classification task.

\section{Method}
In this paper, we focus on differentiating high and low viability teams through chat transcripts of online team interaction. We operationize our goal as a binary classification task. Specifically, we derive both algorithmic and human-coded features from organizational behavior and behavioral science and apply to them to the interaction data from the teams' chat transcripts. The algorithmic features are drawn from a toolkit of modern natural language processing and computational social science tools; human-coded features are labeled by multiple independent workers on Amazon Mechanical Turk. Using these features, we then seek to classify high or low team viability using machine learning models.

In this section we will describe how the team interaction dataset was processed, how the features were encoded on the interaction data, and how the machine learning algorithms were set up to classify team viability.

\subsection{Data processing}\label{data_processing}
\begin{figure}[tb]
 % \centering

 \includegraphics[width=\linewidth]{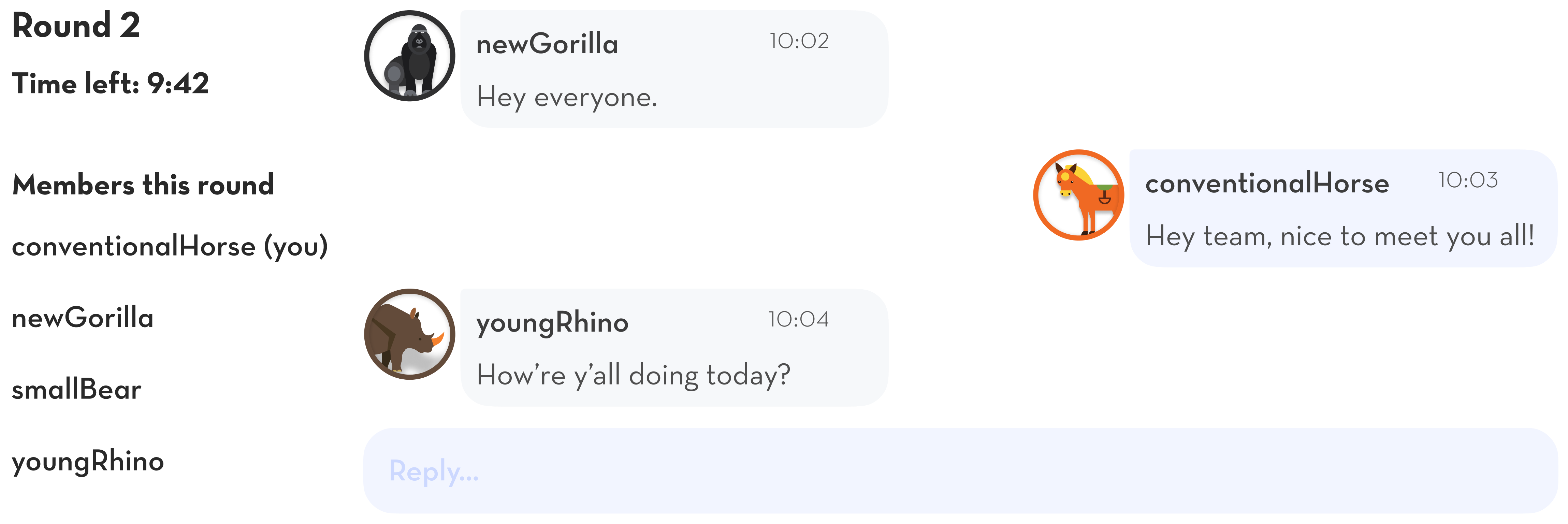}

 \caption{Participants collaborated in a chat room with anonymous usernames. Our dataset of interaction histories consists of all chat messages, who they were sent by, and a timestamp.}
 \label{fig:chat_room}
\end{figure}

We work with data of teams' interactions gathered for a previous study that investigated team fracture and viability~\cite{whiting2019did, whiting2020parallel}. In this dataset, participants recruited from Amazon Mechanical Turk were grouped into teams of 3--8 and remotely collaborated through a synchronous messaging chat room to complete various tasks over several rounds of 10 minutes. The teams are all made up of only US workers who have completed at least 500 tasks on Mechanical Turk (48\% female, and 38 years old on average). The identity of members shown to others was manipulated such that any one group of members would work together 4--6 times without knowing that they had collaborated previously. These data include all of the messages exchanged during the discourse associated with unique pseudonyms and the timestamp since the beginning of each interaction. In these interactions, three different types of tasks from McGrath's task circumplex were used~\cite{straus1999testing}: creative, intellective, and cognitive conflict, although around 80\% of the data utilize the creative task. After completion of each task, team members were asked to complete several surveys, including a 14-item viability instrument on a 5 point Likert scale~\cite{cooperstein2017initial}.

As just mentioned, teams worked together multiple times in succession without knowing they were in fact interacting with the same collaborators~\cite{whiting2020parallel,whiting2019did}. We achieved this effect through a two directional pseudonym masking technique designed to reset team dynamics between rounds --- to participants, it appeared that they were starting a fresh collaboration with new team members. This was validated through an in-sample manipulation check to ensure that collaborators were truly not aware that they were working together repeatedly. Additionally, among the repeated interactions for each team, there was also one pair of interactions where the team members were knowingly \textit{reconvened} with the same group by using consistent pseudonyms. In the prior experiments, this treatment was used to understand path dependence of viability in teams. In this study, we will leave aside data from these repeated teams for every part of our analysis which does not explicitly require reconvened teams, in order to avoid comparing teams with different lengths of interaction. We do take advantage of this unique aspect of the data to specifically make claims about the properties of viability signals in reconvening teams relating to~RQ\ref{resques:howLong}. As far as we know, this is the only chat dataset affording comparative analysis across independent rounds and reconvening teams, which is why we selected it for this paper.

\begin{figure}[tb]
 \includegraphics[width=0.7\linewidth]{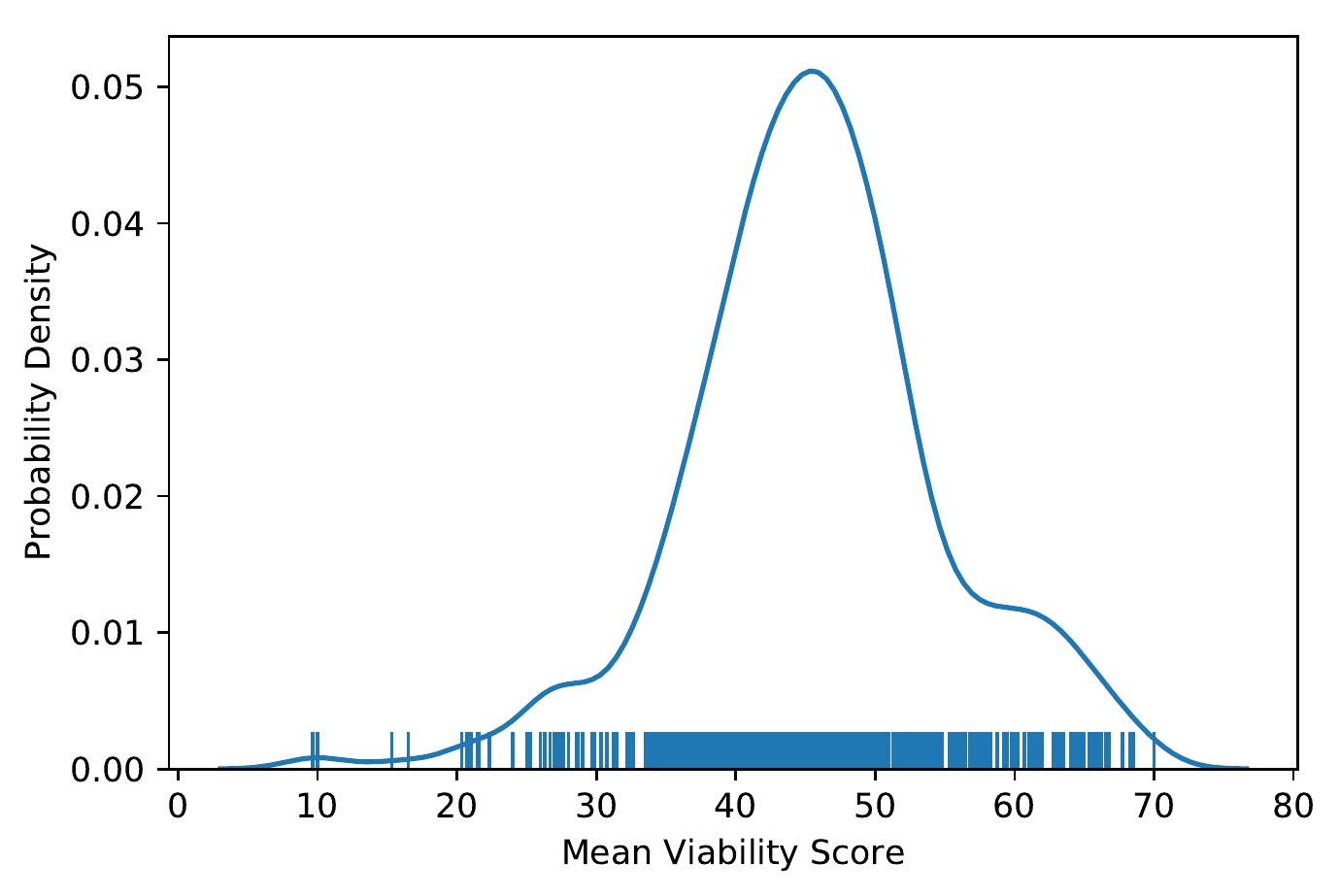}
 \caption{Distribution of average team viability scores across all teams in our dataset.}
 \label{fig:viability}
\end{figure}

In total, we analyzed 669 independent chat interactions of teams, aggregating to a total of 40,024 messages. For each team, we summed each collaborator's 14 viability scale items into a single individual score with a range of 14--70 and calculated the mean of all of the individual scores to determine a team viability score. Figure \ref{fig:viability} illustrates the overall distribution of team viability: the median viability score of all teams is 45.5, the 10th percentile is 34.95, and the 90th percentile is 58.25.

\subsection{Feature Collection}
\subsubsection{Computationally-derived Features}
To study characteristic differences between high and low viability teams, we first derive features through algorithms via the series of chat messages in a multi-person team interaction. These features can be divided into four categories: work pattern, semantics, topic, and word choice, as motivated by prior literature discussed in Section~\ref{sec:antecedents}. Because there are anywhere from 3 to 8 people in a group, we first calculate the features on the level of each individual team member from the entire chat interaction and consequently produce final team-level statistics of that feature (mean, standard deviation, minimum, maximum) that are used to model the chat discourse.

\begin{table}[hbt!]
 \caption{Computationally-derived features associated with work pattern, semantics, topic, and word choice. Some features are multiplied by four, because the minimum, maximum, mean, and standard deviation were calculated for each. 
 }
 \label{tbl:algfeatures}
 \begin{tabular}
 {>{\RaggedRight}p{1in}>{\RaggedRight}p{0.75in}>{\RaggedRight}p{3in}}
 Feature Group & Number & Specific Features \\ 
 \toprule
 Work Pattern & 2$\times$4 & Number of messages per person, number of words per person\\
 Semantics & 2$\times$4 & Subjectivity, polarity \\
 Topic & 1 & TF-IDF cosine similarity \\
 Word Choice & 25 & LIWC (anger, anxiety, sadness, articles, conjunctions, prepositions, inclusive, exclusive, 
 quantifiers, certainty, discrepancies, negation, tentativeness, first person singular, first person plural,
 second person, indefinite pronouns, adverbs, social), argue, reference, readability \\
 
\end{tabular}
\end{table}

\begin{table}[hbt!]
\caption{Computationally-derived features drawn from previous literature.} %\jdz{which of these features fall into which categories?}}
\label{tbl:computation}
% \begin{tabular}{>{\RaggedRight}p{1.7in}>{\RaggedRight}p{3.4in}}

\small
\begin{tabular}{>{\RaggedRight}p{2.3in}>{\RaggedRight}p{2.7in}}
Feature & Measurement Method \\ \toprule
Average message count, Talkative member message count, Laconic member message count, Variation in message counts, Average word count, Talkative member word count, Laconic member word count, Variation in word count~\cite{niculae2016conversational} & Statistical measures on team members' chat messages \\
\addlinespace[0.1cm]
Average polarity score, Highest polarity individual score, Least polarity individual score, Variation in polarity scores, Average subjectivity score, Most subjective individual score, Least subjective individual score, Variation in subjectivity scores~\cite{nguyen2015sentiment} & TextBlob sentiment analysis algorithms evaluated on team members' messages \\
\addlinespace[0.1cm]
Average reading level, Highest individual reading level, Lowest individual reading level, Variation in reading levels~\cite{robinson2013predicting} & Dale-Chall readability score algorithm applied to team members' messages \\
\addlinespace[0.1cm]
Anger words, Anxiety words, Sadness words, Articles, Conjunctions, Prepositions, Inclusive language, Exclusive language, Certain language, Disagreement language, Negation language, Tentative language, First person singular, First person plural, Second person, Indefinite pronouns, Adverbs, Relationship-oriented language~\cite{coppersmith2014quantifying,tausczik2010psychological} & LIWC lexical category counts\\
\addlinespace[0.1cm]
Collaborator language similarity~\cite{somasundaran2010recognizing} & Average cosine similarity of TF-IDF vectors based on unigrams and bigrams\\
\addlinespace[0.1cm]
Argument discourse markers~\cite{abbott2011can, allen2014detecting} & Counts of words from \emph{ARGUE} corpus\\
\addlinespace[0.1cm]
Pseudonym use~\cite{shibani2017assessing} & Counts of referring to others by pseudonym in the chat messages\\

\end{tabular}
\end{table}

With work pattern features, we seek to capture higher level behavioral information from how team members are interacting. In task-oriented discussions, well-balanced participation among group members is a key factor present in productive teams~\cite{niculae2016conversational}. We derive two features in this area: number of messages sent and number of words used by each team member. For each person, we count the number of separate messages he/she sent as well as the total number of words he/she used across all messages for the entire chat discourse. 

For semantic features, we applied text-based sentiment analysis as it has been widely used in predicting human behavior as in social media~\cite{nguyen2015sentiment}. Specifically, we derive the polarity (whether language is positive or negative) and subjectivity (whether language is objective or subjective) of group members' messages using the natural language processing library TextBlob.\footnote{https://textblob.readthedocs.io/en/dev/} 

For the topic features, we expanded upon previous work, which show that unigrams and bigrams are particularly effective for disagreement and argument detection~\cite{somasundaran2010recognizing}. Specifically, we create TF-IDF (term frequency-inverse document frequency) vectors based on unigrams and bigrams as a representation of topics in a person's chat messages. Preprocessing of the messages only involved removing accents to normalize to the Unicode standard, and converting all letters to lowercase; neither stop word removal or other token-level analysis like stemming or lemmatization were used.

The average pairwise cosine similarity of these vectors is used as the feature to describe the alignment of topics in different team members' language.

For the word choice features, we applied dictionary-based text analysis, which has often been used to identify signals in language, for example, to study mental health using social media~\cite{coppersmith2014quantifying}. Specifically, we use LIWC (Linguistic Inquiry Word Count), an extensively human validated lexicon, to count the words that individuals used during the chat discourse that fall into different lexical categories~\cite{tausczik2010psychological}. For our task, we reduce the total category set to a subset of those that are cognitive- or function-oriented and those related to affect (see Table \ref{tbl:algfeatures}), culminating to 19 categories. Additionally, because we expect conversational conflict to influence team viability, we include as one feature the total count for the presence of any 17 discourse markers for recognizing disagreement from the \emph{ARGUE} corpus~\cite{abbott2011can,allen2014detecting}. 
 
Prior work on chat-based team dialogues~\cite{shibani2017assessing} motivated us to incorporate the number of references to other team members as another feature. As a higher level representation of word choice, we also calculate the minimum, maximum, mean, and standard deviation of the Dale-Chall readability scores for each members' chat contribution.

\begin{table}[hbt!]
\caption{Human-labeled features drawn from previous literature and the statements used to ask Amazon Mechanical Turk workers to rate them. The mean standard deviation ($\bar{\sigma}$) is for each feature across all included raters and across interactions is also shown.} 
\label{tbl:human_features}
\small
% \begin{tabular}{>{\RaggedRight}p{1.7in}>{\RaggedRight}p{3.4in}}
\begin{tabular}%{>{\RaggedRight}p{2.3in}>{\RaggedRight}p{2.7in}>{\RaggedRight}}
{>{\RaggedRight}p{2.3in}>{\RaggedRight}p{2.7in}>{\RaggedRight}p{0.1in}}
Feature & Survey Question & $\bar{\sigma}$ \\ \toprule
Generating Many Ideas~\cite{constructivedisc} & There were many ideas/suggestions/input generated from this group. & 0.92 \\ \addlinespace[0.1cm]
Reaching a Conclusion~\cite{dionne2010} & The group reached a decision in the end (there was a sense of consensus). & 0.88 \\ \addlinespace[0.1cm]
Progressing Slowly~\cite{beal2003cohesion} & It took a long time for the team to reach a conclusion. & 1.15 \\ \addlinespace[0.1cm]
Thoughtful Response~\cite{mach2010differential, Barrick1998} & The final answer of the team is reasonable and well-thought-out (e.g: a creative answer for ad writing). & 0.90 \\ \addlinespace[0.1cm]
%\cite{dionne2010} \cite{mach2010differential, Barrick1998}\cite{Kriplean2012, Kirman2012}
Social Loafing~\cite{constructivedisc} & At least one team member was not participating as much as the other team members. & 1.15 \\ \addlinespace[0.1cm]
Annoying Collaborators~\cite{Kriplean2012, Kirman2012} & At least one team member was annoying (used all caps, or sarcasm, or was obnoxious, etc). & 1.32\\ \addlinespace[0.1cm]
Frustrated Collaborators~\cite{Kriplean2012, Kirman2012} & At least one team member expressed frustration/confusion. & 1.27\\ \addlinespace[0.1cm]
Positive Interaction~\cite{barsade2002} & There were positive interactions (emojis, lol, great!, etc). & 0.89\\ \addlinespace[0.1cm]
Starting Salutations~\cite{suchan2001, greenberg2007} & The team members greeted each other at the beginning of the interaction. & 0.87\\ \addlinespace[0.1cm]
Ending Salutations~\cite{greenberg2007} & The team members congratulated each other at the end of the interaction. & 0.96\\ \addlinespace[0.1cm]
Sarcastic Interaction~\cite{muresan2016} & There were sarcastic comments. & 1.28\\ \addlinespace[0.1cm]
Agreeable Interaction~\cite{convogoneawry} & The group was agreeable. & 0.89\\ \addlinespace[0.1cm]
Respectful Interaction~\cite{convogoneawry} & The teammates were supportive and respectful as opposed to insulting. & 0.85\\ \addlinespace[0.1cm]
Passive-aggressive Interaction~\cite{civilizedDiscussion} & Team members were passive-aggressive. & 1.28\\ \addlinespace[0.1cm]
Dismissing Collaborators~\cite{duggan2017, bradley2012reaping} & At least one person in the team was dismissed at some point. & 1.35\\ \addlinespace[0.1cm]
Punishing Collaborators~\cite{duggan2017, bradley2012reaping} & At least one person in the team was punished at some point. & 1.35\\ \addlinespace[0.1cm]
Showing Embarrassment~\cite{duggan2017, bradley2012reaping} & At least one person in the team seemed embarrassed at some point. & 1.31\\ \addlinespace[0.1cm]
Political Interaction~\cite{convogoneawry, Kittur2009, duggan2017, catsrule} & This interaction was Political as opposed to Non-political. & 1.35 \\ \addlinespace[0.1cm]
Playful Interaction~\cite{convogoneawry, Kittur2009, duggan2017, catsrule} & The conversation was Playful as opposed to ideologically charged. & 0.96 \\ \addlinespace[0.1cm]
Fact-driven Discussion~\cite{convogoneawry, Kittur2009, duggan2017, catsrule} & Ideas were Fact-based as opposed to Emotion-based. & 0.99\\ \addlinespace[0.1cm]
\end{tabular}
\end{table}

\subsubsection{Human-labeled Features}\label{sec:crowdfeatures}

Furthermore, we hypothesized that certain aspects of teamwork would be discernible to humans but not to algorithms. We have drawn 20 features from organizational behavior (Table~\ref{tbl:human_features}) to be hand-labeled. We used the Amazon Mechanical Turk platform to recruit $366$ human annotators who had completed at least 500 tasks to annotate teams' chats. For a given chat, the human annotators filled out a survey of 20 questions corresponding to the 20 features, answering each question using a 5-point Likert scale.

More specifically, our $366$ human annotators or "reviewers" must have  completed at least $1000$ tasks and must be based in the USA. The recruiting HIT also excluded all reviewers who had participated in the original studies. Workers were paid at a rate of \$15 per hour, in accordance with existing literature~\cite{whiting2019fair}. We then built a platform that displays one chat per annotation task along with instructions given to the original participants in the task. With all pseudo-names of the participants replaced by a label of "Person\_<number>" (e.g. Person1, Person2, ...) format, the reviewers were asked to rate their perception of the chat interaction on questions measuring selected features on a 5-point Likert scale and to predict whether a team would fracture. We included an attention check question asking for the number of participants in the chat and excluded responses that were not consistent with the truth from the analysis. The order of questions were randomized, and reviewers were allowed to participate in the task multiple times.

At least 3 reviewers annotated each chat, and the median response was selected as the feature value for analysis. The average standard deviation of scores across all raters and chats for a given feature was $1.08$ prior to filtering. Responses that were more than 1.5 standard deviation away from the mean result were excluded prior to calculating the median. Mean standard deviation values for each feature across all interactions are shown in Table~\ref{tbl:human_features}.

\subsection{Machine learning models}
We operationize our goal through a binary classification task to differentiate high and low viability teams. For our analysis, we divide all teams into two groups --- either high viability or low viability --- based on their viability score. Specifically, we set the cutoff viability score as a varying decision threshold, and classify all teams with viability score lower than the cutoff score as low viability teams and teams with viability score above the cutoff as high viability teams. For instance, if we set the cutoff threshold as 45.5 (the median viability score in our dataset), we divide all teams into two equal sized groups, where those teams with viability score above the median set as the high viability sample and those teams with viability lower than the median as the low viability sample. As such, we treat predicting team viability as a binary classification problem because we are interested in distinguishing teams with relatively higher viability from teams with relatively lower viability, rather than predicting the particular measure of viability.

We implemented 5 widely-used machine learning classification models from Python's scikit-learn library~\cite{scikit-learn}, i.e., logistic regression, support vector classification (SVC), random forest classifier, multi-layer perceptron (MLP) classifier, and gradient boosting classifier, to train and classify low vs high team viability based on our derived features. We used the area under the curve of the receiver operating characteristics curve (AUC ROC) as a measure of prediction performance, since we formulated the viability differentiation as a binary classification problem and AUC is insensitive to an unbalanced testing set. Note that most experiments are done via 5-fold stratified cross validation, and each feature in the training and test sets are standardized.
\section{Results}

\begin{table*}[b]
    \centering
    \begin{tabular}{cccccccccc}
    % \toprule
     & \multicolumn{9}{c}{\small{Data Partition Percentile}} \\
    \cmidrule[\heavyrulewidth](r){2-10}
    \textbf{Prediction Model} & \textbf{10} & 20 & 30 & 40 & \textbf{50} & 60 & 70 & 80 & \textbf{90} \\
    \hline
    Logistic Regression & \textbf{0.74} & 0.72 & 0.69 & 0.72 & \textbf{0.75} & 0.74 & 0.78 & 0.81 & \textbf{0.92}\\
    SVC & \textbf{0.63} & 0.70 & 0.67 & 0.69 & \textbf{0.76} & 0.75 & 0.79 & 0.85 & \textbf{0.92}\\
    Random Forest & \textbf{0.66} & 0.72 & 0.69 & 0.72 & \textbf{0.73} & 0.75 & 0.79 & 0.86 & \textbf{0.93}\\
    MLP Classifier & \textbf{0.66} & 0.68 & 0.66 & 0.68 & \textbf{0.71} & 0.71 & 0.74 & 0.80 & \textbf{0.90}\\
    Gradient Boosting Classifier & \textbf{0.67} & 0.70 & 0.67 & 0.72 & \textbf{0.75} & 0.73 & 0.76 & 0.86 & \textbf{0.93}\\
    
    \bottomrule
    \end{tabular}
    \caption{Viability classification performance (measured by ROC AUC) with different division of positive and negative samples using all features.}
    \label{tab:combined_percent}
    % \vspace*{-4mm}
\end{table*}

%\subsection{Different divisions of positive and negative samples}
\subsection{Differentiating team viability under different split}

To answer~RQ\ref{resques:canWe}, we evaluated the five models across nine different trials where we set the viability decision threshold at every ten percentile points, (e.g 10th percentile, 20th percentile, ... , 90th percentile). The models are able to differentiate higher from lower viability teams (Table \ref{tab:combined_percent}).  %suggests that team members' behaviors throughout the interaction have signals relevant to the team's viability. 
Perhaps more importantly, our results suggest that there exist general mechanisms behind team viability vs. team interaction, independent of the team member composition.

When we consider the case of classifying a team's viability relative to the median split, the best AUC score of the five models is 0.76. More interestingly, we find that as the data partition becomes such that we label fewer samples as high viability teams and focus on the teams with relatively higher viability scores, the classification accuracy becomes higher. For instance, when we label the top decile teams as high viability teams and designate the remaining 90\% as low viability teams, we reach over 0.9 AUC, indicating there are salient differences between teams above the 90th percentile and the rest.

In the feature set we are using, in order to capture a wide variety of team interaction behaviors, some features were automatically developed by computational means, and those that could not be, due to the limitations of current algorithms, were labeled by humans. For the practicality of creating a tool for differentiating team viability, we further performed a preliminary ablative analysis of our features. We repeat the previous procedure separately for the computationally derived and human labeled features, which is shown in Table~\ref{tab:alg_percent} and Table~\ref{tab:crowd_percent}, respectively.

\begin{table*}[tb]
     \centering
     \begin{tabular}{cccccccccc}
     & \multicolumn{9}{c}{\small{Data Partition Percentile}} \\
     \cmidrule[\heavyrulewidth](r){2-10}
     % \toprule
    \textbf{Prediction Model} & \textbf{10} & 20 & 30 & 40 & \textbf{50} & 60 & 70 & 80 & \textbf{90} \\
     \hline
     Logistic Regression & \textbf{0.77} & 0.76 & 0.73 & 0.75 & \textbf{0.76} & 0.76 & 0.78 & 0.82 & \textbf{0.94}\\
     SVC & \textbf{0.63} & 0.74 & 0.72 & 0.73 & \textbf{0.76} & 0.75 & 0.79 & 0.84 & \textbf{0.93}\\
     Random Forest & \textbf{0.70} & 0.70 & 0.69 & 0.73 & \textbf{0.72} & 0.73 & 0.77 & 0.84 & \textbf{0.92}\\
     MLP Classifier &	\textbf{0.69} & 0.71 & 0.71 & 0.71 & \textbf{0.73} & 0.74 & 0.75 & 0.82 & \textbf{0.92}\\
     Gradient Boosting Classifier & \textbf{0.66} & 0.69 & 0.67 & 0.73 & \textbf{0.76} & 0.73 & 0.78 & 0.85 & \textbf{0.92}\\
    
     \bottomrule
     \end{tabular}
     \caption{Viability classification performance (measured by ROC AUC) with different division of positive and negative samples by computationally derived features. }
     \label{tab:alg_percent}
     % \vspace*{-4mm}
\end{table*}

 \begin{table*}[tb]
    \centering
     \begin{tabular}{cccccccccc}
     % \toprule
      & \multicolumn{9}{c}{\small{Data Partition Percentile}} \\
     \cmidrule[\heavyrulewidth](r){2-10}
     \textbf{Prediction Model} & \textbf{10} & 20 & 30 & 40 & \textbf{50} & 60 & 70 & 80 & \textbf{90} \\
     \hline
     Logistic Regression & \textbf{0.56} & 0.56 & 0.53 & 0.54 & \textbf{0.61} & 0.60 & 0.63 & 0.69 & \textbf{0.71}\\
     SVC & \textbf{0.46} & 0.44 & 0.47 & 0.51 & \textbf{0.58} & 0.60 & 0.63 & 0.68 & \textbf{0.66}\\
     Random Forest & \textbf{0.58} & 0.50 & 0.65 & 0.55 & \textbf{0.52} & 0.56 & 0.60 & 0.65 & \textbf{0.68}\\
     MLP Classifier &	\textbf{0.55} & 0.52 & 0.51 & 0.53 & \textbf{0.55} & 0.59 & 0.63 & 0.67 & \textbf{0.70}\\
     Gradient Boosting Classifier & \textbf{0.57} & 0.48 & 0.53 & 0.55 & \textbf{0.5} & 0.56 & 0.59 & 0.67 & \textbf{0.70}\\
    
     \bottomrule
     \end{tabular}
     \caption{Viability classification performance (measured by ROC AUC) with different division of positive and negative samples by human labeled features.}
     \label{tab:crowd_percent}
     % \vspace*{-4mm}
 \end{table*}

\begin{figure}[]
    \centering
    \includegraphics[width=0.85\linewidth]{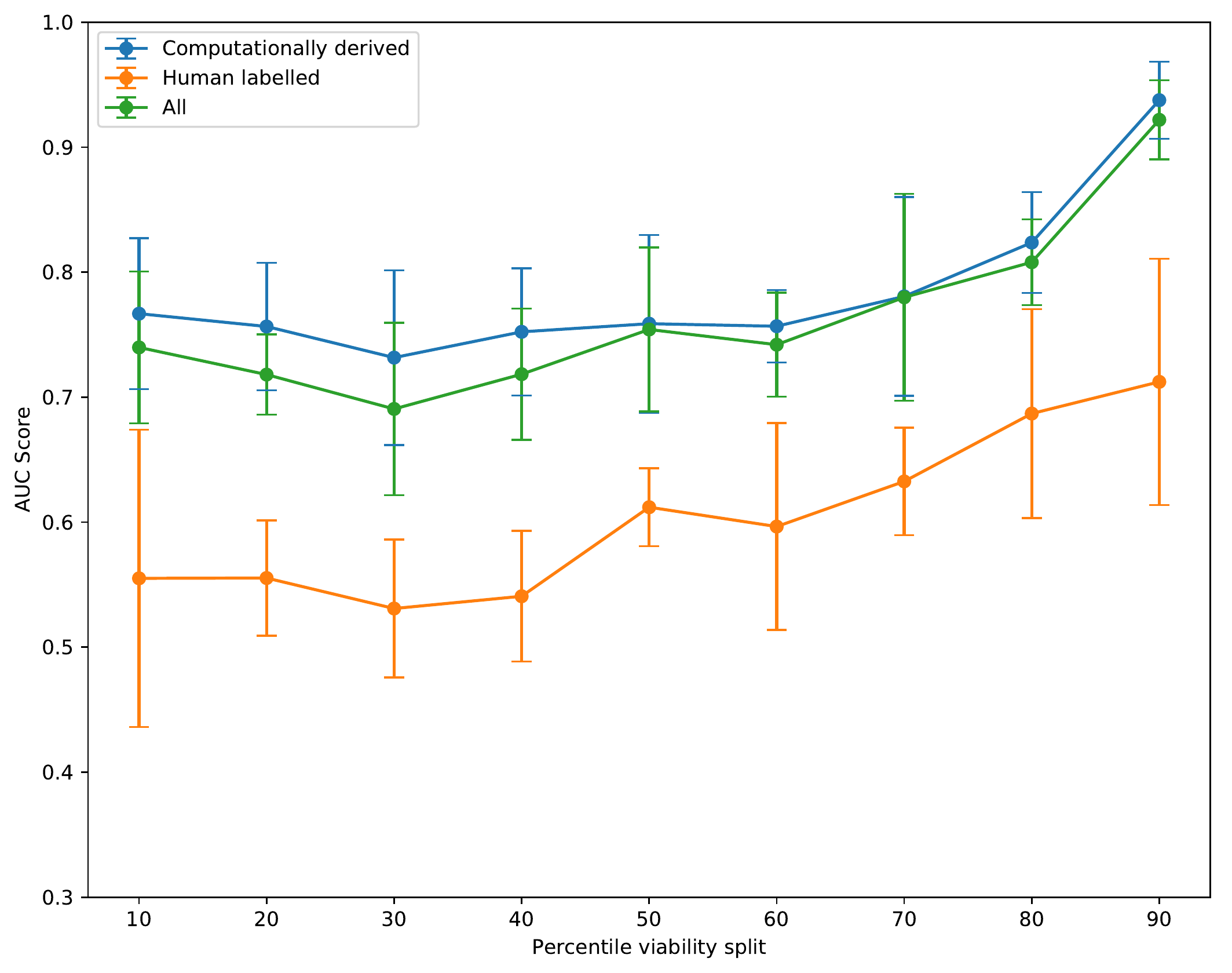}
 
    \caption{Performance of the logistic regression model in AUC score with computationally derived features, human labeled features, and all features across different viability percentile thresholds. Errors bars show the 95\% confidence interval across 5 folds. }
    % \msb{This graph is a bit eye-bleeding to look at. Just pick the best model and graph it; we don't need to see all of them. Your goal is to show that it goes up as you focus on splitting just the higher viability teams.}
    % \msb{what do the error bars in this prediction task mean?}
    \label{fig:percentiles_graph}
\end{figure}

We see a noticeable difference in the performance of the two methods, but regardless, we witness a similar pattern of increasing accuracy as the threshold for dividing the data increases, which can be seen in the right-side plot of Figure \ref{fig:percentiles_graph}. Across all decision thresholds, the classification accuracy of the computationally derived features is at least as good as the full features set, having an AUC of 0.76 of at the median split and 0.94 at the 90th percentile split as seen in Table \ref{tab:alg_percent}. Using our human labeled features alone results in only an AUC of 0.61 at the median split and 0.71 at the 90th percentile split as seen in Table \ref{tab:crowd_percent}. This discrepancy could suggest that 1) there can be significant signal overlap between computationally derived and human labeled features, or 2) human labeled features may be noisier than computationally derived features. This result provides evidence that computationally derived features alone may be suited to further study. 

Finally, to rule out the possibility that team performance may be a confounder, we carried out a robustness check by measuring the correlation between team performance (as measured by actual task performance, e.g. click through rate of advertisement text in the creative task) and team viability. In our dataset, 175 teams have team performance information. We observe weak correlation between the two ($R^2=0.10$). As another test, we add team performance as a feature to the classification model and compare it with the classification model without team performance on these 175 teams. We find no change in classification AUC: under 10\% and 50\% splits, the AUC is stable at 0.78, and under 90\% split, the AUC is stable at 0.96 under both conditions, which further verifies that team viability cannot be explained by team performance.

For the rest of our analysis, we will focus on using our logistic regression model. It achieves the best results in most viability splits and as a model, it is computationally simpler and more interpretable than the other models we tested. Additionally, we report other performance metrics specifically for logistic regression under various conditions of decision threshold and used features in Table~\ref{tab:more_metrics}.

\begin{table}[]

\small
\begin{tabular}{cccccc}
% \specialrule{1pt}{1pt}{1pt}
% \hline
\toprule
\textbf{Features} & \textbf{\shortstack{Partition\\Percentile}}  & \textbf{F1 Score} & \textbf{Precision} &\textbf{Recall} & \textbf{ROC AUC}\\
\hline
\multirow{3}{*}{\shortstack{Computationally\\Derived}} & 10  & 0.10 & 0.13 & 0.07 & 0.77\\
                             & 50  & 0.67 & 0.70 & 0.64 & 0.76\\
                             & 90  & 0.57 & 0.69 & 0.52 & 0.94\\
\hline
\multirow{3}{*}{Human Labelled}& 10  & 0.03 & 0.20 & 0.02 & 0.56 \\
                             & 50  & 0.57 & 0.59 & 0.56 & 0.61\\
                             & 90  & 0.08 & 0.50 & 0.05 & 0.71\\
\hline
\multirow{3}{*}{All Features}         & 10  & 0.20 & 0.32 & 0.15 & 0.74 \\
                             & 50  & 0.66 & 0.67 & 0.64 & 0.75 \\
                             & 90  & 0.56 & 0.62 & 0.52 & 0.92 \\
\bottomrule
% \hline
% \specialrule{1pt}{1pt}{1pt}
\end{tabular}
\caption{Additional metrics of viability classification for Logistic Regression when using a decision threshold at the 10th, 50th, and 90th percentile of viability score. Note that AUC is a better metric to evaluate viability classification performance in our case since the metric based on ranking, while F1 score, precision and recall are based on threshold to divide positive and negative labels. The high AUC for our model indicates the ability of our model to correctly rank high viability team ahead of low viability team, while the relatively low F1 score only reflects the cutoff threshold is not ideal.}
\label{tab:more_metrics}
\vspace{-2em}

\end{table}

\subsection{Feature Analysis}

\begin{figure}[tb]
    \centering
    \includegraphics[width=0.9\linewidth]{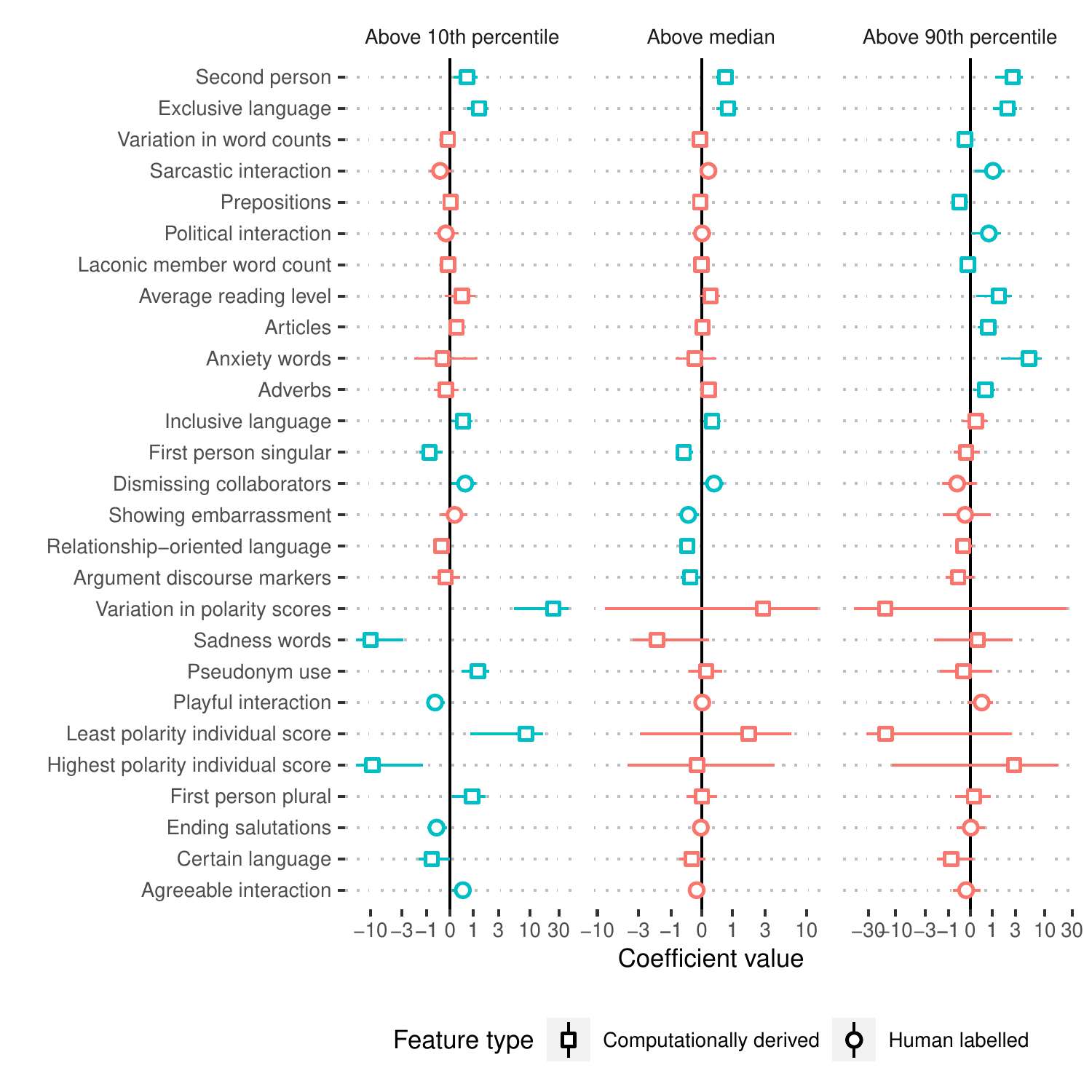}
    \caption{Logistic regression coefficients for features. Features with confidence interval distinguishable from zero (significant) are shown in blue, while features with confidence interval overlapping zero (insignificant) are shown in red. Features are ordered from most to least useful in classification, and they are clustered by viability split. Values are shown on a hyperbolic arcsine scale to cope with variance.}
    \label{fig:combined_feature_analysis}
\end{figure}

To answer~RQ\ref{resques:whichFeatures}, we proceed to investigate the role of different features in predicting team viability. Specifically, we run penalized logistic regression with L1 regularization (Lasso regression) for our model using threshold scores of 34.95 (bottom decile split), 45.5 (median split), and  58.25 (top decile split), and examine the coefficients of the features to understand whether certain features contribute positively or negatively to team viability. We choose penalized logistic regression since it helps prevent collinearity and sparsity with the effect that the coefficients' weights are moved to the most predictive features.

We present coefficients and their confidence intervals for all the significant features in Fig.~\ref{fig:combined_feature_analysis}. As can be observed, 

two features stand out as significant features across all three splits. Use of exclusive language (as defined by LIWC), such as `but', `either', `except', and addressing each other in second person (e.g. `you') are significantly and positively correlated with the teams with relatively higher viability across all three splits, which support findings that suggest teams whose members actively respond to and engage with the ideas of others are more likely to be viable~\cite{behfar2008critical}. In one of the top decile teams, the team members used exclusive language to consider a variety of different viewpoints and arguments when discussing their solution, 

\begin{quote}
\textbf{Person3:} I believe it would be fair to allocate most, if not all of the funds to project 3. Neither of the other two projects are vital, in my opinion.\\
\textbf{Person4:} great idea but sooo many people are hurting \\
\textbf{Person2:} \#2 is the artsy option that we might get donations from the art community or art enthusiasts to help
\end{quote}

On the other other hand, we observe that there are quite a few non-overlapping features that are significant when differentiating high and low viability teams at different thresholds, indicating that bottom decile viability teams, above and below median teams and top decile viability teams demonstrate different interaction behavior.

What distinguishes top decile teams from other teams? The use of exclusive language and second person are very strongly correlated with top decile teams, even more so than the other two splits for which these two features are also positive indicators. Other features such as articles and adverbs support this overarching trend of active engagement with other team members. As such, more active engagement appears to be an extremely important trend for top decile teams. A higher average reading level for the chat interaction is also positively correlated with membership in the top decile, suggesting that top decile teams often have the willingness and ability to communicate complex topics. In addition, the top decile has two interesting indicators which seem unintuitive upon first glance. The human-labeled presence of sarcastic interaction is positively correlated with viability scores above the 90th percentile. Upon further investigation, we found that while sarcasm has a negative connotation generally, the instances of human-labeled sarcasm which were present in top decile team interactions were of the humorous variety. 
\begin{quote}
\textbf{Person2:} 5) it's in the 120 range maybe? anyone live in Death Valley? ;)
\end{quote}
In addition, the computationally-derived feature of Anxiety Words was also a positive indicator of top decile teams, but further investigation reveals that team members were using words which the algorithm identified as anxious in order to talk about their task, e.g., that "veterans were \textit{struggling}". 

What makes a team more likely to be a bottom decile team? The significant features that stand out make intuitive and theoretical sense. More exhibits of sadness, as well as the presence of a highly polarized individual --- an individual who outputs very positive or very negative chat --- are associated with the bottom decile of teams. These findings are consistent with the organizational behavior theory that displays of emotion, especially negative ones, will negatively impact the effectiveness of the team.~\cite{lewis2000leaders}
Interestingly, more use of first person singular terms (e.g., I, me, mine) and less use of first person plural (e.g., we, our, us) and second person words (e.g, you, your) is also associated with a higher chance of being in the bottom decile teams. This trend contributes to our conclusion that more active engagement with other team members correlates with higher viability, and less active engagement with other members correlates with lower viability. Finally, playful interactions were correlated with the bottom decile teams as well. Further investigation shows that overly playful teams had difficulty actually resolving the task at hand, ultimately leading to lower viability scores. For example, one team had a hard time deciding on a slogan for a chair after joking about a pun, 
\begin{quote}
\textbf{Person3:} What about Bar Stool for American's Behind\\
\textbf{Person4:} that's America's Ass! \\
\textbf{Person4:} it would get meme points lol\\
\textbf{Person3:} A Stool from America's Ass \\
...\\
\textbf{Person2:} we have to decide ppl 
\end{quote}
When distinguishing teams above and below the median viability score, the general trends we discover in the more extreme splits hold true. More use of exclusive language and second person pronouns, which directly engage with other team members, are positive indicators of team viability greater than the median. The presence of argument discourse markers, and exhibitions of embarrassment, are correlated with team viability lower than the median. Two unusual features are specifically significant for the median viability split. Dismissing collaborators is a positive indicator for teams above median viability, which further corroborates the overarching trend that directly engaging with other team members --- even when such engagement is active disagreement --- is correlated with higher viability, since direct engagement allows team members to feel that the team actively communicates on issues. Finally, relationship-oriented language is interestingly correlated with teams below the median viability. We note that this LIWC category contained words that were colloquial and could easily be interpreted negatively, e.g. "give advice" and "bro". While these words could be used positively in some situations, we expect that their negative interpretation may be disproportionately more impactful on team viability, while their positive interpretation may have little to no impact.

\subsection{Differentiating team viability over time}
To answer~RQ\ref{resques:howLong}, we further investigate whether our model can be used to classify team viability over time. Specifically, we are curious about whether mechanisms behind team viability are time dependent or generalizable over time, i.e. is a team interaction - viability model learnt from team interactions at a certain stage applicable to classification of team viability at another stage? 

The dataset we use is well-suited to study this research question since in the experiment design, people are asked to engage in multiple rounds of tasks (round 1 to round 4), and for each round they join newly-convened teams with teammates they do not recognize. 

To shed light on this question, we ran a series of ablative experiments where for $j \in [1,4]$, we trained a logistic regression model using team interactions from all rounds except round $j$ and tested the model on round $j$, i.e. can we use interaction patterns mined from rounds other than $j$ to predict team viability at round $j$? Under this setting, we expect to observe rather consistent viability classification performance should the relationship between viability and team dynamics be static, otherwise there will be a significant performance drop since signals from round $j$ are not encoded in the trained model. 

In other words, if the behaviors that are associated with high or low viability are stable, we would expect the model trained on a team's first meeting to still function well on their fourth meeting; if the behaviors are not stable, we would expect the model to have weaker performance when run on a different meeting than its training set. We ran separate experiments for all features, computationally derived features and human labeled features with varying decision thresholds of viability scores. Additionally, we ensure there are equal numbers of training/testing samples (300 training, 60 testing) across trials to provide a fair comparison.

\begin{figure}[tb]
    \centering
    \includegraphics[width=0.9\linewidth]{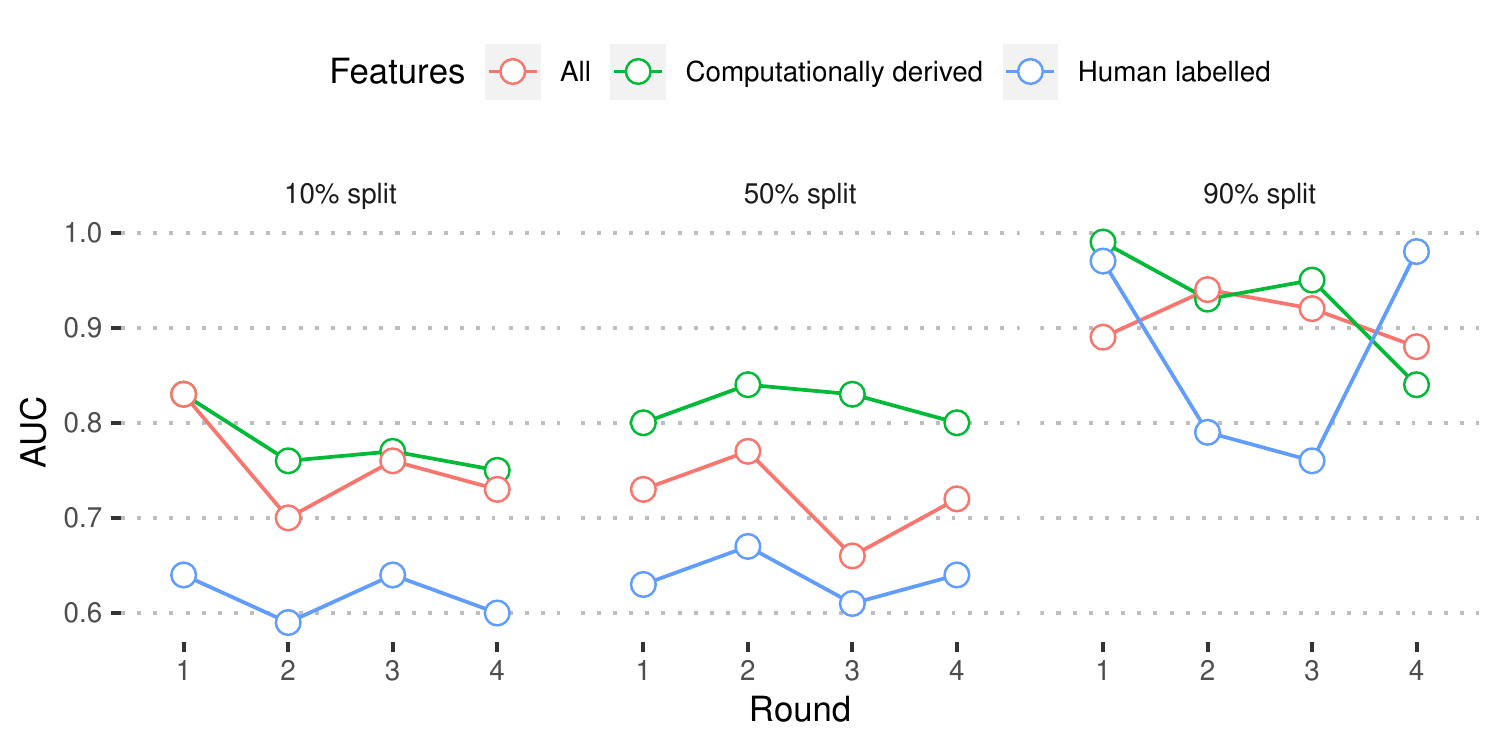}
    \caption{Predictive accuracy across rounds appears to not show any major trend. Notably, no confidence intervals are provided because these could not be calculated with cross validation --- these results are exact given the data studied.}
    \label{fig:round_analysis}
\end{figure}

We present the results in Fig~\ref{fig:round_analysis}. As can be observed, in the case of the 10\% and 50\% split, there is not much deviation in the model's performance between different rounds used to train the model whether all features, computationally derived or human labeled features are used. Classification performance is stable around AUC of 0.8 for all features and algorithm derived features while around AUC of 0.65 for human labeled features. For the 90\% viability split, we do observe a large drop from AUC close to 1.0 to AUC 0.8 for human labeled features at rounds 2 and 3 and for computationally derived features at round 4, yet there is no clear monotonic trend between performance and testing round indicating the variation is likely due to noise. Moreover, classification performance for all features remains relatively stable around 0.9 under the 90\% split. Thus we conclude that our proposed model is rather robust regardless of the length of time team members have engaged in the work, which provides evidence that the relationship between viability and team interaction is static over time. 

\subsection{Differentiating Initial and Reconvened Team Viability}
\begin{figure}[tb]
    \centering
    \includegraphics[width=0.9\linewidth]{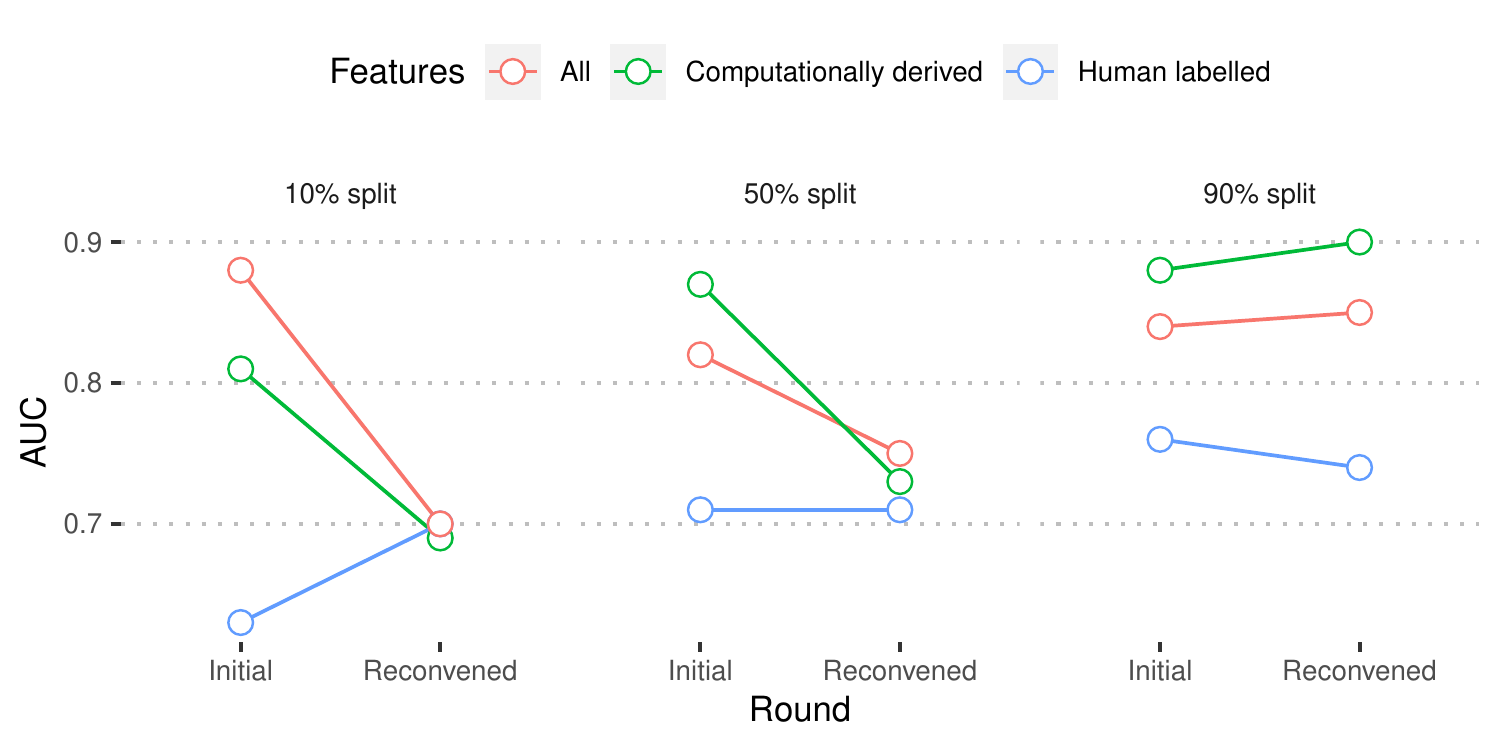}
    \caption{Predictive accuracy in reconvened rounds sees a substantial dip for algorithmic features in lower viability split cases, while human features appear less affected. Notably, no confidence intervals are provided because these could not be calculated with cross validation --- these results are exact given the data studied.}
    \label{fig:reconvening_analysis}
\end{figure}

As mentioned in Section~\ref{data_processing}, the dataset we used has a unique property of reconvened teams~\cite{whiting2019did,whiting2020parallel} in addition to the independent rounds we have just discussed. In reconvened rounds, team members were aware that they were collaborating with the same team members for more than one separate round. These reconvened teams have not appeared in our training or testing sets so far to avoid any potential confounds, however, here we are interested to investigate whether our model will perform differently under the reconvened scenario, so as to answer~RQ\ref{resques:howLong} from another perspective. Specifically, we trained the viability classification model using all the masked interactions (300 interactions), and test on unmasked initial interaction (133 rounds) and reconvened interaction (132 rounds), for the three feature groups under the 10\%, 50\% and 90\% division of viability. The result is reported in Figure~\ref{fig:reconvening_analysis}.

We find that when using all features and computationally derived features, classifying team viability for reconvened interactions achieves much lower performance compared to classifying team viability for the initial interactions when we use the 10\% and 50\% division: for example, for the 10\% split, computational features achieve 0.81 AUC when classifying the initial interaction but only 0.69 AUC when reconvened. Given prior analysis where we have shown that patterns behind team viability is rather consistent over time for 10\% and 50\% division, we conclude that the result indicates that computationally derived features are not so consistent at differentiating team viability in the reconvened condition, which may be a result of the model failing to capture signals embedded in the fact that team members know they are working with the same group of people for the reconvened rounds. Meanwhile, at the 90th percentile split, the viability classification performance remains relatively stable, indicating that the top decile viable teams have behavior patterns that remain consistent despite knowing they are working with the same group.  

Interestingly, we find that for human labeled features, the viability classification performances for the initial and reconvened interactions are identical or even increase, indicating that human labeled features are perhaps more robust at classifying reconvened interactions.

\subsection{Differentiating viability based on different amounts of interaction}
Finally, to answer~RQ\ref{resques:howShort}, we performed an additional analysis using our computationally derived features to test how little of the chat discourse is necessary to effectively differentiate teams' viabilities. Because we will need to consider smaller windows of the chat interactions, we can only apply the features that were derived computationally; the method with which we produced human labeled features was only performed once at the end of the whole chat interaction. We expect that the group interactions are dynamic and may change over the course of completing the task. For this study, since the chat histories we are analyzing are about 10 minutes long, we consider windows of data at 10 second increments, totaling to 60 trials. We start by calculating our features on the first ten seconds and evaluating the performance of our model; next, we take the data within the first 20 second window, calculate features, and evaluate the AUC score; this continues until the full 10 minutes of data is used. All trials were performed only with our Logistic Regression model with three different viability threshold splits: 10\%, 50\%, and 90\%.

\begin{figure}[tb]
    \centering
    \includegraphics[width=\linewidth]{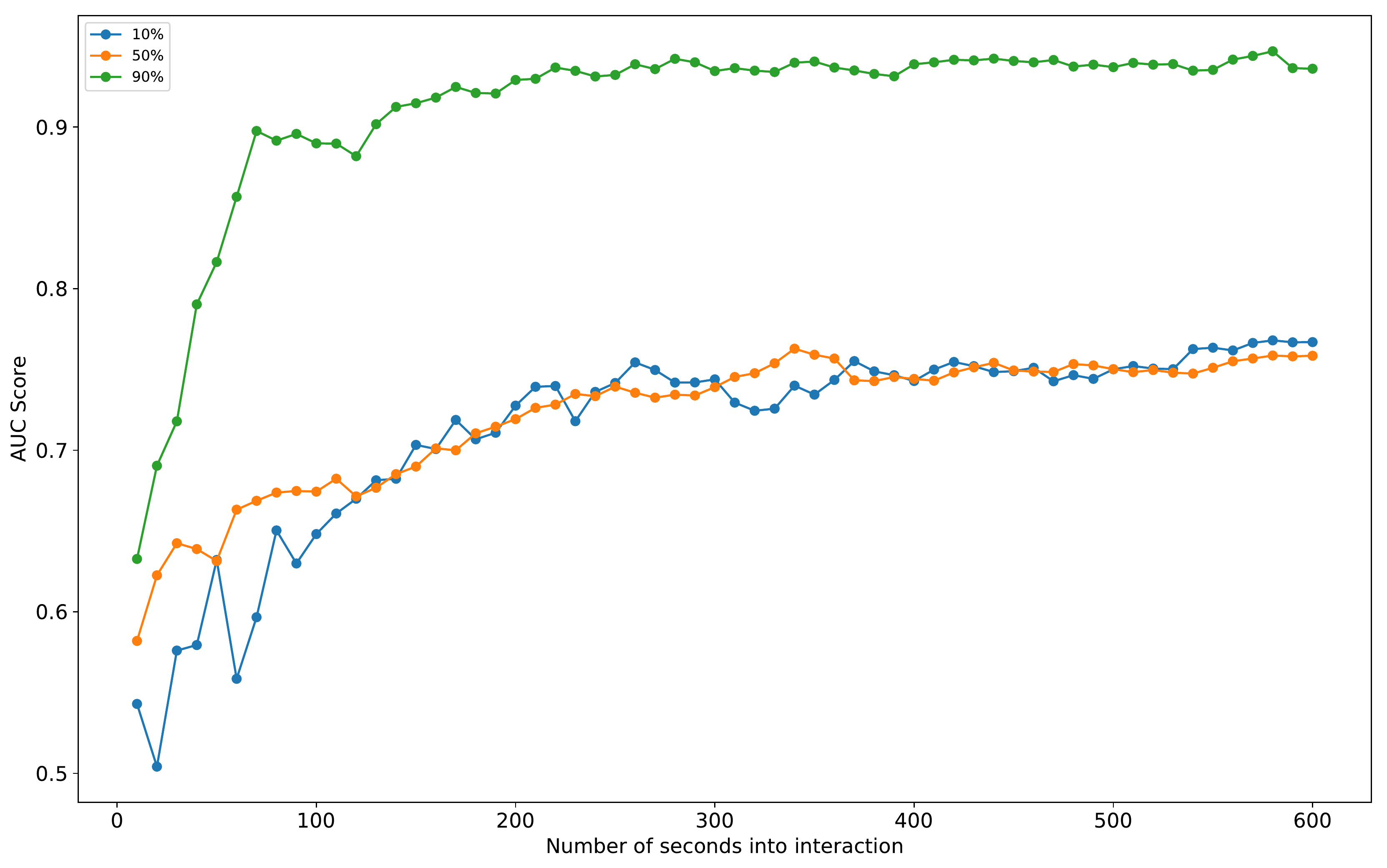}
    \caption{Results of training the Logistic Regression model with data from different numbers of seconds of the chat interaction when low v. high viability is split at the 10, 50, and 90 percentile viability.}
    \label{fig:interaction_portion}
\end{figure}

In Figure \ref{fig:interaction_portion}, we see that when using the 10th and 50th percentile splits, the rate of improvement is gradual, and reaches consistent performance when at least 200 seconds of chat data is used. In these two cases, our approach obtains 95\% of the full accuracy with at least 35\% of the entire interaction. The performance trend when analyzing the top decile teams (90\% split) is notably different; as the number of seconds of data increases, the AUC score rapidly increases and converges to steady performance as soon as when the training procedure considers at about 70 seconds of chat data. With only 12\% of the chat interaction, we achieve 95\% of the full accuracy when differentiating the top decile teams. This could suggest that the teams in the top decile of team viability exhibit some behavior earlier on that strongly contributes to predicting their final team viability. Regardless of the viability threshold used to divide the teams, we see that our model can consistently predict low v. high team viability much earlier in the process than at the full 10 minute mark. 

\begin{figure}[tb]
    \centering
    \includegraphics[width=0.75\linewidth]{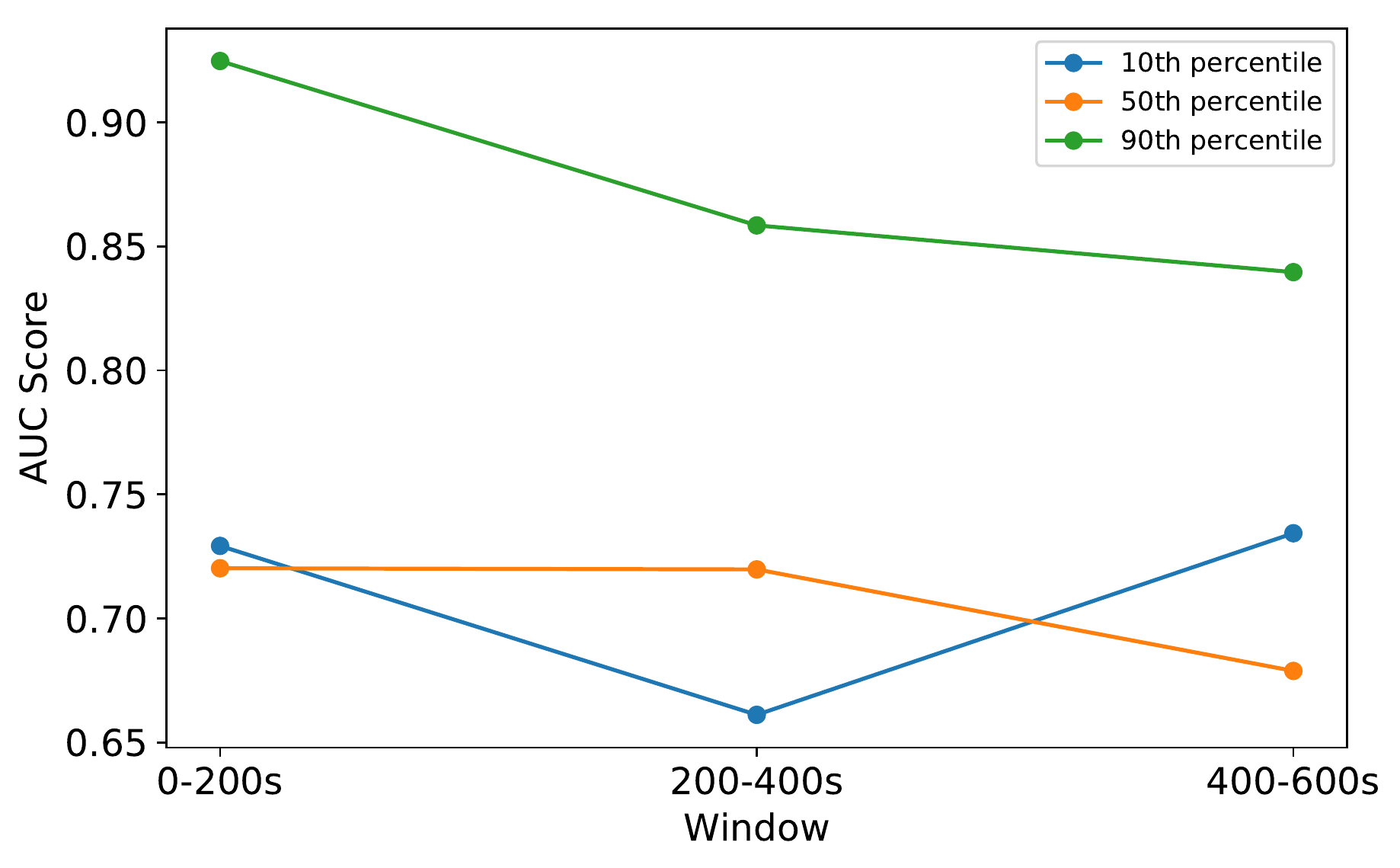}
    \caption{Performance of Logistic Regression model trained on disjoint windows of 200 seconds at different percentile splits (10th, 50th, 90th).}
    \label{fig:thin_slice_win}
\end{figure}

To understand why this small window of chat data early in the collaboration can be predictive of the team viability by the end of the interaction, we compare the performance of consecutive disjoint windows of data to capture behaviors at different times in the interaction. Because the model achieved 95\% of the full accuracy in about 200 seconds for the 10th and median viability split, we divide the ten minute interaction into three 200 second windows. Whereas the previous thin slicing analysis was done by considering cumulative windows of data, this time, none of the windows of data for each trial share any data. Looking at Figure~\ref{fig:thin_slice_win}, we see that the performance for all three viability splits remains relatively consistent over the three windows. This suggests that a small slice of data in the initial stages of the interaction is effective because by 200 seconds, team members start to exhibit consistent signals of behaviors that influence viability. Latter windows in the middle and end of the interaction continue to have the presence of these behaviors, hence the relatively consistent performance scores. 

\section{Discussion}
\subsection{Research Implications}

The strong performance of our models in classifying teams by viability suggests that team viability largely is driven by underlying mechanisms that are common across different teams and repeated interactions of the same team. In particular, we find that work patterns and the language usage, which can be computationally derived, are effective in differentiating high and low viability teams with high accuracy of AUC $= 0.76$ when equally splitting the data, and AUC $\geq 0.9$ when differentiating the top decile of high viability teams from the rest. 
Furthermore, the considerably high accuracy for classifying the teams with the top decile viability score and that only two features are shared as being significant across all decision thresholds of viability score suggest that the behavior of the most viable teams are notably different from that of the rest of the teams, with lowest to moderate viability. The relative decrease in the performance when classifying relative to the median or 10th percentile viability score implies that differentiating teams with an average level of viability and those with the lowest viability is more challenging. While all teams with the highest viability in our dataset are viable for the same reasons, indicated by the common signals of active contribution, equal division of labor, more accessible, objective, and positive language, less viable teams appear to reach that state for differing reasons, not as consistently as the former. 

We also note that a majority of our significant features in all three classification viability thresholds (10th, 50th, 90th) were computationally derived, so they can be evaluated without exposing the interaction text to another party or person and remain accurate given a fraction of each interaction. On the other hand, human-labelled features were more robust against changes in context, suggested by the consistency in differentiating team viability in the initial interactions and the reconvened rounds. Although our features are not exhaustive by any means, we suggest that automated features, which offer a consistent comparison of all teams, identify distinctive characteristics of teams depending on their viability, while the human-labelled features reflect human judgment that can detect nuanced signals of team dynamics, such as sarcastic humor, more efficiently.

\subsection{Design Implications} 
Our work also has several design implications for building computer-supported collaboration platforms. For instance, in online crowd-sourced collaboration systems, we show that it is possible to differentiate high and low viability teams with even a few instances of member interaction, providing opportunities for timely intervention if necessary. Designers could potentially leverage our detection pipelines to help teams monitor their viability and build intervention tools that respond to warning signals to promote a successful collaboration experience. To illustrate this, if the system recognizes signals associated with low team viability (e.g., overuse of negation, absence of second person pronouns), the system can remind the teams to take steps for positive behavior change.

Such systems can help improve the sustainability and longevity of teams in the workplace. Today, many teams rely on human leadership (i.e. a manager or a supervisor) to resolve team viability issues. However, such human figures can only detect and intervene after an incident has already occurred. 
Often times, after an incident like a heated dispute occurs, a team's dynamics are beyond remedy. Meanwhile, the aforementioned systems can detect team viability issues in real time, allowing leadership to intervene before a major conflict erupts. This system affords teams a chance to resolve any growing issues and stay together before an irrevocable episode materializes. 

The features used by these systems can also shed light on why a team's viability is suffering and recommend tailored intervention depending on the factors for low viability. For example, if a team's low viability can be explained by highly unequal participation, managers can subsequently address this specific issue and improve the team's work patterns. On the other hand, if low viability of a team can be attributed to overuse of negation, a nudging reminder to be more positive may be more effective. It is important to note that a combination of behaviors may be affecting a team's viability. Exploring different combinations may allow systems to learn the behavior patterns of a team and make more personalized recommendations for intervention. Furthermore, understanding how more viable teams behave can help provide areas for improvement for teams that are not necessarily considered not viable but simply want to improve how they work together. To illustrate, behavior associated with highest viability teams can be referenced when a team establishes a team norm.

Despite the previously mentioned beneficial use cases, these systems may also result in undesirable outcomes. To illustrate, a faulty prediction of the system can lead to the deterioration of viability going unnoticed until a major episode occurs or costly and unnecessary interventions. A more concerning implication of an automated system assessing viability is the possibility of the system adopting biases against certain communities based on distinctive language patterns, such as  culturally-dependent colloquialism or accents. If the system is more sensitive to the language of certain groups of people, it may systematically and wrongly flag them as being problematic. Such an algorithmic bias would lead to discriminatory practices and inequitable treatment based on one's cultural identity, which is both detrimental to team viability and unacceptable. Finally, the system may be hijacked for malicious or inappropriate uses. For example, the sheer amount of data that would be collected regarding individuals' dialogue behavior raises concern for protecting one's privacy against surveillance. Furthermore, the system could also be inappropriately applied to predict characteristics other than viability (e.g. performance), leading to flawed evaluations and perceptions of teams. 

These issues motivate a variety of design principles and guidelines for usage in order to avoid unnecessary cost, inefficacy, injustices, and malicious usage. First, signals used to assess viability should be made public and should be readily audited. Further, indications of viability should not be made for individuals and attempts to measure individual viability potential should be avoided. Since viability is inherently a team property, extrapolation of assessment to individuals will be both flimsy and potentially damaging. Although it is tempting to develop recommendations with high precision, the trade-off between maximizing precision and minimizing potential abuse of targeted recommendations should be considered seriously. To resolve these issues in our own work, we are making available a version of our algorithm in which the features and their coefficients are transparently reported, and additionally, where a suitable amount of interaction material will be required to make assessments. Further discussion of ideas
to respect privacy and ensure algorithmic fairness should be continued throughout the process of design, implementation, and deployment.

\subsection{Limitations \& Future Work}

Considering the methods with which our dataset was gathered, there are a few limitations to the generalizability of the insights we draw from our analysis. First, the teams we analyze are relatively small in size and cannot represent the diversity and complexity in team roles and hierarchies in large organizations. Second, there may be important differences between remote collaboration and collocated work (even though the two forms demonstrate considerable characteristic similarities~\cite{cao2020your,chen2020understanding}), particularly around finding common ground~\cite{olson2000distance}. Prior research suggests that conflict of all types (task, affective, and process) will be detrimental to the performance of remote teams but may not have the same effects on traditional teams~\cite{hinds2003out}. Thus, our approach may not fully capture interactions in collocated teams. Third, the chat interactions themselves are short in duration (10 mins for each round) and involve well-structured tasks. Coordination and problem defining in more complex, open-ended tasks may follow different dynamics than what participants did in our study, which could affect a team's viability. 

In future work, it would be interesting to extend this approach to real world cases like software engineering tasks, which is a prolonged collaboration and involves complex interactions between people and the work itself. This work focuses on remote synchronous collaboration, but whether the same insights apply to face-to-face interactions or what differences there may be are both useful areas to explore. Also, due to the size of the collected dataset, we are unable to do fine-grained analysis comparing our model under different task categories. In the future, it would be interesting to extend our analysis to different task categories.

Nevertheless, we believe our derived insights do generalize and can be of great value to understand team viability under a remote collaboration/interaction setting (which is becoming increasingly important as more collaborations are shifted online after the COVID-19 pandemic), especially for short-term team interactions where members do not know each other's identities. Such flash interactions indeed happen often in online communities, e.g., in online forums, discussions under social media posts, and news reviews, and are also becoming common in crowdsourcing tasks. As such interactions are essentially similar in property to our collected data, we argue similar patterns are very likely to hold. Our findings and implications on team viability can help better understand user engagement in these settings.

Another possible line of analysis is delving into the cultural context of the teams, the human raters, and the computationally judged features. In the current work, all of the data are collected from US Amazon Mechanical Turk workers with moderate experience, and this sampling is likely to influence aspects of our result. Notably, the researchers attempted to code chats using the same prompts as the human raters did and found no notable difference in accuracy or inter rater reliability. Further, no stratification of the human rating data implicate sub populations within our analysis who systematically performed better. With this said, we expect these results may differ when other populations or cultural backgrounds are engaged. Finally, it would be useful to include other factors of interpersonal communications (e.g., linguistic accommodation, between individual personality/dissimilarity, etc.) as future work to gain a deeper understanding of team viability.

\section{Conclusion}
Recognizing the importance of team viability --- rather than performance --- in effective collaboration, we study signals of team viability in text interaction data of remote teams. Our machine learning models using features motivated by previous literature can reliably differentiate teams that belong to the top decile, above median, and the bottom decile of viability scores. Further investigation of the models suggest that while the use of exclusive language and second person pronouns are associated with higher viability teams across thresholds, other behaviors are unique to distinguishing the top decile, such as higher average reading level, and the bottom decile, such as sadness and displays of extreme emotions. Furthermore, we demonstrate that the relationship between viability and team interaction is consistent over time, yet the pattern differs by de novo teams and reconvened teams. Finally, we find that signals from the initial moments of interaction are sufficient for successful prediction of the whole interaction, which suggests that viability can be evaluated and acted upon early. The strength of our models using automated features indicate opportunities for real time assessment of viability, which lends further space for intervention.

\begin{acks}
We are grateful for the efforts of numerous research assistants, Amazon Mechanical Turk worker participants and Upwork research engineers --- without them we would not have been able to collect this data. We thank Daniel A. McFarland for helpful discussions and suggestions. We also thank the associate editor and reviewers for their valuable comments. The data collection was supported by the Stanford Data Science Initiative, RISE Thailand Consortium, the Hasso Plattner Design Thinking Research Program, the Office of Naval Research (N00014-16-1-2894) and a National Science Foundation award IIS-1351131.
\end{acks}

\bibliographystyle{ACM-Reference-Format}
\bibliography{refer}


\begin{thebibliography}{72}

%%% the Web address.

\ifx \showCODEN    \undefined \def \showCODEN     #1{\unskip}     \fi
\ifx \showDOI      \undefined \def \showDOI       #1{#1}\fi
\ifx \showISBNx    \undefined \def \showISBNx     #1{\unskip}     \fi
\ifx \showISBNxiii \undefined \def \showISBNxiii  #1{\unskip}     \fi
\ifx \showISSN     \undefined \def \showISSN      #1{\unskip}     \fi
\ifx \showLCCN     \undefined \def \showLCCN      #1{\unskip}     \fi
\ifx \shownote     \undefined \def \shownote      #1{#1}          \fi
\ifx \showarticletitle \undefined \def \showarticletitle #1{#1}   \fi
\ifx \showURL      \undefined \def \showURL       {\relax}        \fi
% invisible to TeX
\providecommand\bibfield[2]{#2}
\providecommand\bibinfo[2]{#2}
\providecommand\natexlab[1]{#1}
\providecommand\showeprint[2][]{arXiv:#2}

\bibitem[\protect\citeauthoryear{Abbott, Walker, Anand, Fox~Tree, Bowmani, and
  King}{Abbott et~al\mbox{.}}{2011}]%
        {abbott2011can}
\bibfield{author}{\bibinfo{person}{Rob Abbott}, \bibinfo{person}{Marilyn
  Walker}, \bibinfo{person}{Pranav Anand}, \bibinfo{person}{Jean~E Fox~Tree},
  \bibinfo{person}{Robeson Bowmani}, {and} \bibinfo{person}{Joseph King}.}
  \bibinfo{year}{2011}\natexlab{}.
\newblock \showarticletitle{How can you say such things?!?: Recognizing
  disagreement in informal political argument}. In
  \bibinfo{booktitle}{\emph{Proceedings of the Workshop on Languages in Social
  Media}}. Association for Computational Linguistics, \bibinfo{pages}{2--11}.
\newblock


\bibitem[\protect\citeauthoryear{Allen, Carenini, and Ng}{Allen
  et~al\mbox{.}}{2014}]%
        {allen2014detecting}
\bibfield{author}{\bibinfo{person}{Kelsey Allen}, \bibinfo{person}{Giuseppe
  Carenini}, {and} \bibinfo{person}{Raymond Ng}.}
  \bibinfo{year}{2014}\natexlab{}.
\newblock \showarticletitle{Detecting disagreement in conversations using
  pseudo-monologic rhetorical structure}. In
  \bibinfo{booktitle}{\emph{Proceedings of the 2014 Conference on Empirical
  Methods in Natural Language Processing (EMNLP)}}.
  \bibinfo{pages}{1169--1180}.
\newblock


\bibitem[\protect\citeauthoryear{Anand, Walker, Abbott, Tree, Bowmani, and
  Minor}{Anand et~al\mbox{.}}{2011}]%
        {catsrule}
\bibfield{author}{\bibinfo{person}{Pranav Anand}, \bibinfo{person}{Marilyn
  Walker}, \bibinfo{person}{Rob Abbott}, \bibinfo{person}{Jean E.~Fox Tree},
  \bibinfo{person}{Robeson Bowmani}, {and} \bibinfo{person}{Michael Minor}.}
  \bibinfo{year}{2011}\natexlab{}.
\newblock \showarticletitle{Cats Rule and Dogs Drool!: Classifying Stance in
  Online Debate}. In \bibinfo{booktitle}{\emph{Proceedings of the 2Nd Workshop
  on Computational Approaches to Subjectivity and Sentiment Analysis}}
  \emph{(\bibinfo{series}{WASSA '11})}. \bibinfo{publisher}{Association for
  Computational Linguistics}, \bibinfo{address}{Stroudsburg, PA, USA},
  \bibinfo{pages}{1--9}.
\newblock
\showISBNx{9781937284060}
\urldef\tempurl%
\url{http://dl.acm.org/citation.cfm?id=2107653.2107654}
\showURL{%
\tempurl}


\bibitem[\protect\citeauthoryear{Aube and Rousseau}{Aube and Rousseau}{2005}]%
        {aube2005team}
\bibfield{author}{\bibinfo{person}{Caroline Aube} {and}
  \bibinfo{person}{Vincent Rousseau}.} \bibinfo{year}{2005}\natexlab{}.
\newblock \showarticletitle{Team goal commitment and team effectiveness: the
  role of task interdependence and supportive behaviors.}
\newblock \bibinfo{journal}{\emph{Group Dynamics: Theory, Research, and
  Practice}} \bibinfo{volume}{9}, \bibinfo{number}{3} (\bibinfo{year}{2005}),
  \bibinfo{pages}{189}.
\newblock


\bibitem[\protect\citeauthoryear{Barrick, Stewart, Neubert, and Mount}{Barrick
  et~al\mbox{.}}{1998a}]%
        {barrick1998relating}
\bibfield{author}{\bibinfo{person}{Murray~R Barrick}, \bibinfo{person}{Greg~L
  Stewart}, \bibinfo{person}{Mitchell~J Neubert}, {and}
  \bibinfo{person}{Michael~K Mount}.} \bibinfo{year}{1998}\natexlab{a}.
\newblock \showarticletitle{Relating member ability and personality to
  work-team processes and team effectiveness.}
\newblock \bibinfo{journal}{\emph{Journal of applied psychology}}
  \bibinfo{volume}{83}, \bibinfo{number}{3} (\bibinfo{year}{1998}),
  \bibinfo{pages}{377}.
\newblock


\bibitem[\protect\citeauthoryear{Barrick, Stewart, Neubert, and Mount}{Barrick
  et~al\mbox{.}}{1998b}]%
        {Barrick1998}
\bibfield{author}{\bibinfo{person}{Murray~R Barrick}, \bibinfo{person}{Greg~L
  Stewart}, \bibinfo{person}{Mitchell~J Neubert}, {and}
  \bibinfo{person}{Michael~K Mount}.} \bibinfo{year}{1998}\natexlab{b}.
\newblock \showarticletitle{Relating member ability and personality to
  work-team processes and team effectiveness.}
\newblock \bibinfo{journal}{\emph{Journal of applied psychology}}
  \bibinfo{volume}{83}, \bibinfo{number}{3} (\bibinfo{year}{1998}),
  \bibinfo{pages}{377}.
\newblock


\bibitem[\protect\citeauthoryear{Barsade}{Barsade}{2002}]%
        {barsade2002}
\bibfield{author}{\bibinfo{person}{Sigal~G. Barsade}.}
  \bibinfo{year}{2002}\natexlab{}.
\newblock \showarticletitle{The Ripple Effect: Emotional Contagion and Its
  Influence on Group Behavior}.
\newblock \bibinfo{journal}{\emph{Administrative Science Quarterly}}
  (\bibinfo{year}{2002}), \bibinfo{pages}{644--675}.
\newblock
\urldef\tempurl%
\url{https://doi.org/10.2307/3094912}
\showURL{%
\tempurl}


\bibitem[\protect\citeauthoryear{Beal, Cohen, Burke, and McLendon}{Beal
  et~al\mbox{.}}{2003}]%
        {beal2003cohesion}
\bibfield{author}{\bibinfo{person}{Daniel~J Beal}, \bibinfo{person}{Robin~R
  Cohen}, \bibinfo{person}{Michael~J Burke}, {and} \bibinfo{person}{Christy~L
  McLendon}.} \bibinfo{year}{2003}\natexlab{}.
\newblock \showarticletitle{Cohesion and performance in groups: A meta-analytic
  clarification of construct relations.}
\newblock \bibinfo{journal}{\emph{Journal of applied psychology}}
  \bibinfo{volume}{88}, \bibinfo{number}{6} (\bibinfo{year}{2003}),
  \bibinfo{pages}{989}.
\newblock


\bibitem[\protect\citeauthoryear{Behfar, Peterson, Mannix, and Trochim}{Behfar
  et~al\mbox{.}}{2008}]%
        {behfar2008critical}
\bibfield{author}{\bibinfo{person}{Kristin~J Behfar},
  \bibinfo{person}{Randall~S Peterson}, \bibinfo{person}{Elizabeth~A Mannix},
  {and} \bibinfo{person}{William~MK Trochim}.} \bibinfo{year}{2008}\natexlab{}.
\newblock \showarticletitle{The critical role of conflict resolution in teams:
  A close look at the links between conflict type, conflict management
  strategies, and team outcomes.}
\newblock \bibinfo{journal}{\emph{Journal of applied psychology}}
  \bibinfo{volume}{93}, \bibinfo{number}{1} (\bibinfo{year}{2008}),
  \bibinfo{pages}{170}.
\newblock


\bibitem[\protect\citeauthoryear{Bell and Marentette}{Bell and
  Marentette}{2011a}]%
        {bell2011}
\bibfield{author}{\bibinfo{person}{Suzanne Bell} {and} \bibinfo{person}{Brian
  Marentette}.} \bibinfo{year}{2011}\natexlab{a}.
\newblock \showarticletitle{Team viability for long-term and ongoing
  organizational teams}.
\newblock \bibinfo{journal}{\emph{Organizational Psychology Review}}
  \bibinfo{volume}{1} (\bibinfo{date}{11} \bibinfo{year}{2011}),
  \bibinfo{pages}{275--292}.
\newblock
\urldef\tempurl%
\url{https://doi.org/10.1177/2041386611405876}
\showDOI{\tempurl}


\bibitem[\protect\citeauthoryear{Bell}{Bell}{2007}]%
        {bell2007deep}
\bibfield{author}{\bibinfo{person}{Suzanne~T Bell}.}
  \bibinfo{year}{2007}\natexlab{}.
\newblock \showarticletitle{Deep-level composition variables as predictors of
  team performance: a meta-analysis.}
\newblock \bibinfo{journal}{\emph{Journal of applied psychology}}
  \bibinfo{volume}{92}, \bibinfo{number}{3} (\bibinfo{year}{2007}),
  \bibinfo{pages}{595}.
\newblock


\bibitem[\protect\citeauthoryear{Bell and Marentette}{Bell and
  Marentette}{2011b}]%
        {bell2011team}
\bibfield{author}{\bibinfo{person}{Suzanne~T Bell} {and}
  \bibinfo{person}{Brian~J Marentette}.} \bibinfo{year}{2011}\natexlab{b}.
\newblock \showarticletitle{Team viability for long-term and ongoing
  organizational teams}.
\newblock \bibinfo{journal}{\emph{Organizational Psychology Review}}
  \bibinfo{volume}{1}, \bibinfo{number}{4} (\bibinfo{year}{2011}),
  \bibinfo{pages}{275--292}.
\newblock


\bibitem[\protect\citeauthoryear{Bradley, Postlethwaite, Klotz, Hamdani, and
  Brown}{Bradley et~al\mbox{.}}{2012}]%
        {bradley2012reaping}
\bibfield{author}{\bibinfo{person}{Bret~H Bradley}, \bibinfo{person}{Bennett~E
  Postlethwaite}, \bibinfo{person}{Anthony~C Klotz}, \bibinfo{person}{Maria~R
  Hamdani}, {and} \bibinfo{person}{Kenneth~G Brown}.}
  \bibinfo{year}{2012}\natexlab{}.
\newblock \showarticletitle{Reaping the benefits of task conflict in teams: The
  critical role of team psychological safety climate.}
\newblock \bibinfo{journal}{\emph{Journal of Applied Psychology}}
  \bibinfo{volume}{97}, \bibinfo{number}{1} (\bibinfo{year}{2012}),
  \bibinfo{pages}{151}.
\newblock


\bibitem[\protect\citeauthoryear{Cao, Chen, Xu, Wang, Xu, Zhang, and Li}{Cao
  et~al\mbox{.}}{2020}]%
        {cao2020your}
\bibfield{author}{\bibinfo{person}{Hancheng Cao}, \bibinfo{person}{Zhilong
  Chen}, \bibinfo{person}{Fengli Xu}, \bibinfo{person}{Tao Wang},
  \bibinfo{person}{Yujian Xu}, \bibinfo{person}{Lianglun Zhang}, {and}
  \bibinfo{person}{Yong Li}.} \bibinfo{year}{2020}\natexlab{}.
\newblock \showarticletitle{When Your Friends Become Sellers: An Empirical
  Study of Social Commerce Site Beidian}. In
  \bibinfo{booktitle}{\emph{Proceedings of the International AAAI Conference on
  Web and Social Media}}, Vol.~\bibinfo{volume}{14}. \bibinfo{pages}{83--94}.
\newblock


\bibitem[\protect\citeauthoryear{Carr{\`e}re and Gottman}{Carr{\`e}re and
  Gottman}{1999}]%
        {carrere1999predicting}
\bibfield{author}{\bibinfo{person}{Sybil Carr{\`e}re} {and}
  \bibinfo{person}{John~Mordechai Gottman}.} \bibinfo{year}{1999}\natexlab{}.
\newblock \showarticletitle{Predicting divorce among newlyweds from the first
  three minutes of a marital conflict discussion}.
\newblock \bibinfo{journal}{\emph{Family process}} \bibinfo{volume}{38},
  \bibinfo{number}{3} (\bibinfo{year}{1999}), \bibinfo{pages}{293--301}.
\newblock


\bibitem[\protect\citeauthoryear{Cascio and Montealegre}{Cascio and
  Montealegre}{2016}]%
        {Cascio2016}
\bibfield{author}{\bibinfo{person}{Wayne Cascio} {and} \bibinfo{person}{Ramiro
  Montealegre}.} \bibinfo{year}{2016}\natexlab{}.
\newblock \showarticletitle{How Technology Is Changing Work and Organizations}.
\newblock \bibinfo{journal}{\emph{Annual Review of Organizational Psychology
  and Organizational Behavior}}  \bibinfo{volume}{3} (\bibinfo{date}{03}
  \bibinfo{year}{2016}), \bibinfo{pages}{349--375}.
\newblock
\urldef\tempurl%
\url{https://doi.org/10.1146/annurev-orgpsych-041015-062352}
\showDOI{\tempurl}


\bibitem[\protect\citeauthoryear{Chen, Cao, Xu, Cheng, Wang, and Li}{Chen
  et~al\mbox{.}}{2020}]%
        {chen2020understanding}
\bibfield{author}{\bibinfo{person}{Zhilong Chen}, \bibinfo{person}{Hancheng
  Cao}, \bibinfo{person}{Fengli Xu}, \bibinfo{person}{Mengjie Cheng},
  \bibinfo{person}{Tao Wang}, {and} \bibinfo{person}{Yong Li}.}
  \bibinfo{year}{2020}\natexlab{}.
\newblock \showarticletitle{Understanding the Role of Intermediaries in Online
  Social E-commerce: An Exploratory Study of Beidian}.
\newblock \bibinfo{journal}{\emph{Proceedings of the ACM on Human-Computer
  Interaction}} \bibinfo{volume}{4}, \bibinfo{number}{CSCW2}
  (\bibinfo{year}{2020}), \bibinfo{pages}{1--24}.
\newblock


\bibitem[\protect\citeauthoryear{Cooperstein}{Cooperstein}{2017}]%
        {cooperstein2017initial}
\bibfield{author}{\bibinfo{person}{Jessica~Nicole Cooperstein}.}
  \bibinfo{year}{2017}\natexlab{}.
\newblock \showarticletitle{Initial Development of a Team Viability Measure}.
\newblock  (\bibinfo{year}{2017}).
\newblock


\bibitem[\protect\citeauthoryear{Coppersmith, Dredze, and Harman}{Coppersmith
  et~al\mbox{.}}{2014}]%
        {coppersmith2014quantifying}
\bibfield{author}{\bibinfo{person}{Glen Coppersmith}, \bibinfo{person}{Mark
  Dredze}, {and} \bibinfo{person}{Craig Harman}.}
  \bibinfo{year}{2014}\natexlab{}.
\newblock \showarticletitle{Quantifying mental health signals in Twitter}. In
  \bibinfo{booktitle}{\emph{Proceedings of the workshop on computational
  linguistics and clinical psychology: From linguistic signal to clinical
  reality}}. \bibinfo{pages}{51--60}.
\newblock


\bibitem[\protect\citeauthoryear{Daim, Ha, Reutiman, Hughes, Pathak, Bynum, and
  Bhatla}{Daim et~al\mbox{.}}{2012}]%
        {daim2012exploring}
\bibfield{author}{\bibinfo{person}{Tugrul~U Daim}, \bibinfo{person}{Anita Ha},
  \bibinfo{person}{Shawn Reutiman}, \bibinfo{person}{Brennan Hughes},
  \bibinfo{person}{Ujjal Pathak}, \bibinfo{person}{Wayne Bynum}, {and}
  \bibinfo{person}{Ashok Bhatla}.} \bibinfo{year}{2012}\natexlab{}.
\newblock \showarticletitle{Exploring the communication breakdown in global
  virtual teams}.
\newblock \bibinfo{journal}{\emph{International Journal of Project Management}}
  \bibinfo{volume}{30}, \bibinfo{number}{2} (\bibinfo{year}{2012}),
  \bibinfo{pages}{199--212}.
\newblock


\bibitem[\protect\citeauthoryear{Dionne, Sayama, Hao, and Benjamin}{Dionne
  et~al\mbox{.}}{2010}]%
        {dionne2010}
\bibfield{author}{\bibinfo{person}{Shelley~D. Dionne}, \bibinfo{person}{Hiroki
  Sayama}, \bibinfo{person}{Chanyu Hao}, {and} \bibinfo{person}{Benjamin~J.
  Benjamin}.} \bibinfo{year}{2010}\natexlab{}.
\newblock \showarticletitle{The role of leadership in shared mental model
  convergence and team performance improvement: An agent-based computational
  model}.
\newblock  (\bibinfo{year}{2010}), \bibinfo{pages}{1035--1049}.
\newblock
\urldef\tempurl%
\url{https://doi.org/10.1016/j.leaqua.2010.10.007}
\showURL{%
\tempurl}


\bibitem[\protect\citeauthoryear{Discourse}{Discourse}{[n. d.]}]%
        {civilizedDiscussion}
\bibfield{author}{\bibinfo{person}{Discourse}.} \bibinfo{year}{[n.
  d.]}\natexlab{}.
\newblock \showarticletitle{This is a Civilized Place for Public Discussion}.
\newblock  (\bibinfo{year}{[n. d.]}).
\newblock
\urldef\tempurl%
\url{http://bit.ly/1TM8K5x}
\showURL{%
\tempurl}


\bibitem[\protect\citeauthoryear{Duggan}{Duggan}{2017}]%
        {duggan2017}
\bibfield{author}{\bibinfo{person}{Maeve Duggan}.}
  \bibinfo{year}{2017}\natexlab{}.
\newblock \showarticletitle{Online Harassment 2017}.
\newblock \bibinfo{journal}{\emph{Pew Research Center}} (\bibinfo{year}{2017}).
\newblock


\bibitem[\protect\citeauthoryear{Edmondson}{Edmondson}{1999}]%
        {edmondson1999psychological}
\bibfield{author}{\bibinfo{person}{Amy Edmondson}.}
  \bibinfo{year}{1999}\natexlab{}.
\newblock \showarticletitle{Psychological safety and learning behavior in work
  teams}.
\newblock \bibinfo{journal}{\emph{Administrative science quarterly}}
  \bibinfo{volume}{44}, \bibinfo{number}{2} (\bibinfo{year}{1999}),
  \bibinfo{pages}{350--383}.
\newblock


\bibitem[\protect\citeauthoryear{Fussell}{Fussell}{1992}]%
        {FussellKrauss1992}
\bibfield{author}{\bibinfo{person}{R.~M.~Krauss. Fussell, S.~R.}}
  \bibinfo{year}{1992}\natexlab{}.
\newblock \showarticletitle{Coordination of knowledge in communication: Effects
  of speakers’ assumptions about what others know}.
\newblock \bibinfo{journal}{\emph{Journal of Personality and Social
  Psychology}}  \bibinfo{volume}{62} (\bibinfo{year}{1992}).
\newblock


\bibitem[\protect\citeauthoryear{Gottman and Levenson}{Gottman and
  Levenson}{2000}]%
        {gottman2000timing}
\bibfield{author}{\bibinfo{person}{John~Mordechai Gottman} {and}
  \bibinfo{person}{Robert~Wayne Levenson}.} \bibinfo{year}{2000}\natexlab{}.
\newblock \showarticletitle{The timing of divorce: Predicting when a couple
  will divorce over a 14-year period}.
\newblock \bibinfo{journal}{\emph{Journal of Marriage and Family}}
  \bibinfo{volume}{62}, \bibinfo{number}{3} (\bibinfo{year}{2000}),
  \bibinfo{pages}{737--745}.
\newblock


\bibitem[\protect\citeauthoryear{Greenberg, Greenberg, and Antonucci}{Greenberg
  et~al\mbox{.}}{2007}]%
        {greenberg2007}
\bibfield{author}{\bibinfo{person}{Penelope~S. Greenberg},
  \bibinfo{person}{Ralph~H. Greenberg}, {and} \bibinfo{person}{Yvonne~L.
  Antonucci}.} \bibinfo{year}{2007}\natexlab{}.
\newblock \showarticletitle{Creating and sustaining trust in virtual teams}.
\newblock  (\bibinfo{year}{2007}), \bibinfo{pages}{325--333}.
\newblock
\urldef\tempurl%
\url{https://doi.org/10.1016/j.bushor.2007.02.005}
\showURL{%
\tempurl}


\bibitem[\protect\citeauthoryear{Hackman}{Hackman}{1980}]%
        {hackman1980work}
\bibfield{author}{\bibinfo{person}{J~Richard Hackman}.}
  \bibinfo{year}{1980}\natexlab{}.
\newblock \showarticletitle{Work redesign and motivation.}
\newblock \bibinfo{journal}{\emph{Professional Psychology}}
  \bibinfo{volume}{11}, \bibinfo{number}{3} (\bibinfo{year}{1980}),
  \bibinfo{pages}{445}.
\newblock


\bibitem[\protect\citeauthoryear{Hackman}{Hackman}{2011}]%
        {hackman2011collaborative}
\bibfield{author}{\bibinfo{person}{J~Richard Hackman}.}
  \bibinfo{year}{2011}\natexlab{}.
\newblock \bibinfo{booktitle}{\emph{Collaborative intelligence: Using teams to
  solve hard problems}}.
\newblock \bibinfo{publisher}{Berrett-Koehler Publishers}.
\newblock


\bibitem[\protect\citeauthoryear{Hinds}{Hinds}{2003}]%
        {HindsPamela2003}
\bibfield{author}{\bibinfo{person}{Pamela Hinds}.}
  \bibinfo{year}{2003}\natexlab{}.
\newblock \showarticletitle{Out of Sight, Out of Sync: Understanding Conflict
  in Distributed Teams}.
\newblock \bibinfo{journal}{\emph{Organization Science}}  \bibinfo{volume}{14}
  (\bibinfo{date}{11} \bibinfo{year}{2003}), \bibinfo{pages}{615--632}.
\newblock
\urldef\tempurl%
\url{https://doi.org/10.1287/orsc.14.6.615.24872}
\showDOI{\tempurl}


\bibitem[\protect\citeauthoryear{Hinds and Bailey}{Hinds and Bailey}{2003}]%
        {hinds2003out}
\bibfield{author}{\bibinfo{person}{Pamela~J Hinds} {and}
  \bibinfo{person}{Diane~E Bailey}.} \bibinfo{year}{2003}\natexlab{}.
\newblock \showarticletitle{Out of sight, out of sync: Understanding conflict
  in distributed teams}.
\newblock \bibinfo{journal}{\emph{Organization science}} \bibinfo{volume}{14},
  \bibinfo{number}{6} (\bibinfo{year}{2003}), \bibinfo{pages}{615--632}.
\newblock


\bibitem[\protect\citeauthoryear{Hjalmarson and Strandmark}{Hjalmarson and
  Strandmark}{2012}]%
        {hjalmarson2012forming}
\bibfield{author}{\bibinfo{person}{Helene~V Hjalmarson} {and}
  \bibinfo{person}{Margaretha Strandmark}.} \bibinfo{year}{2012}\natexlab{}.
\newblock \showarticletitle{Forming a learning culture to promote fracture
  prevention activities}.
\newblock \bibinfo{journal}{\emph{Health Education}} (\bibinfo{year}{2012}).
\newblock


\bibitem[\protect\citeauthoryear{Jung}{Jung}{2016}]%
        {jung2016coupling}
\bibfield{author}{\bibinfo{person}{Malte~F Jung}.}
  \bibinfo{year}{2016}\natexlab{}.
\newblock \showarticletitle{Coupling interactions and performance: Predicting
  team performance from thin slices of conflict}.
\newblock \bibinfo{journal}{\emph{ACM Transactions on Computer-Human
  Interaction (TOCHI)}} \bibinfo{volume}{23}, \bibinfo{number}{3}
  (\bibinfo{year}{2016}), \bibinfo{pages}{1--32}.
\newblock


\bibitem[\protect\citeauthoryear{Kirman, Lineham, and Lawson}{Kirman
  et~al\mbox{.}}{2012}]%
        {Kirman2012}
\bibfield{author}{\bibinfo{person}{Ben Kirman}, \bibinfo{person}{Conor
  Lineham}, {and} \bibinfo{person}{Shaun Lawson}.}
  \bibinfo{year}{2012}\natexlab{}.
\newblock \showarticletitle{Exploring mischief and mayhem in social computing
  or: how we learned to stop worrying and love the trolls}. In
  \bibinfo{booktitle}{\emph{Extended Abstracts on Human Factors in Computing
  Systems}} \emph{(\bibinfo{series}{CHI '12})}. \bibinfo{publisher}{ACM},
  \bibinfo{address}{New York, NY, USA}, \bibinfo{pages}{121--130}.
\newblock
\showISBNx{978-1-4503-1016-1}
\urldef\tempurl%
\url{https://doi.org/10.1145/2212776.2212790}
\showDOI{\tempurl}


\bibitem[\protect\citeauthoryear{Kittur, Suh, and Chi}{Kittur
  et~al\mbox{.}}{2009}]%
        {Kittur2009}
\bibfield{author}{\bibinfo{person}{Aniket Kittur}, \bibinfo{person}{Bongwon
  Suh}, {and} \bibinfo{person}{Ed~H. Chi}.} \bibinfo{year}{2009}\natexlab{}.
\newblock \showarticletitle{What\'s in Wikipedia? Mapping Topics and Conflict
  Using Socially Annotated Category Structure}. In
  \bibinfo{booktitle}{\emph{Proceedings of the SIGCHI Conference on Human
  Factors in Computing Systems}} \emph{(\bibinfo{series}{CHI '09})}.
  \bibinfo{publisher}{ACM}, \bibinfo{address}{New York, NY, USA},
  \bibinfo{pages}{1509--1512}.
\newblock
\showISBNx{978-1-60558-246-7}
\urldef\tempurl%
\url{https://doi.org/10.1145/1518701.1518930}
\showDOI{\tempurl}


\bibitem[\protect\citeauthoryear{Kozlowski}{Kozlowski}{1998}]%
        {kozlowski1998training}
\bibfield{author}{\bibinfo{person}{Steve~WJ Kozlowski}.}
  \bibinfo{year}{1998}\natexlab{}.
\newblock \showarticletitle{Training and developing adaptive teams: Theory,
  principles, and research.}
\newblock  (\bibinfo{year}{1998}).
\newblock


\bibitem[\protect\citeauthoryear{Kozlowski and Bell}{Kozlowski and
  Bell}{2007}]%
        {kozlowski2007team}
\bibfield{author}{\bibinfo{person}{Steve~WJ Kozlowski} {and}
  \bibinfo{person}{Bradford~S Bell}.} \bibinfo{year}{2007}\natexlab{}.
\newblock \showarticletitle{Team learning, development, and adaptation}.
\newblock In \bibinfo{booktitle}{\emph{Work group learning}}.
  \bibinfo{publisher}{Psychology Press}, \bibinfo{pages}{39--68}.
\newblock


\bibitem[\protect\citeauthoryear{Kozlowski, Gully, Nason, and Smith}{Kozlowski
  et~al\mbox{.}}{1999}]%
        {kozlowski1999developing}
\bibfield{author}{\bibinfo{person}{Steve~WJ Kozlowski},
  \bibinfo{person}{Stanley~M Gully}, \bibinfo{person}{Earl~R Nason}, {and}
  \bibinfo{person}{Eleanor~M Smith}.} \bibinfo{year}{1999}\natexlab{}.
\newblock \showarticletitle{Developing adaptive teams: A theory of compilation
  and performance across levels and time}.
\newblock \bibinfo{journal}{\emph{Pulakos (Eds.), The changing nature of work
  performance: Implications for staffing, personnel actions, and development}}
  \bibinfo{volume}{240} (\bibinfo{year}{1999}), \bibinfo{pages}{292}.
\newblock


\bibitem[\protect\citeauthoryear{Kriplean, Toomim, Morgan, Borning, and
  Ko}{Kriplean et~al\mbox{.}}{2012}]%
        {Kriplean2012}
\bibfield{author}{\bibinfo{person}{Travis Kriplean}, \bibinfo{person}{Michael
  Toomim}, \bibinfo{person}{Jonathan Morgan}, \bibinfo{person}{Alan Borning},
  {and} \bibinfo{person}{Andrew Ko}.} \bibinfo{year}{2012}\natexlab{}.
\newblock \showarticletitle{Is This What You Meant? Promoting Listening on the
  Web with Reflect}. In \bibinfo{booktitle}{\emph{Proceedings of the SIGCHI
  Conference on Human Factors in Computing Systems}}
  \emph{(\bibinfo{series}{CHI '12})}. \bibinfo{publisher}{ACM},
  \bibinfo{address}{New York, NY, USA}, \bibinfo{pages}{1559--1568}.
\newblock
\showISBNx{978-1-4503-1015-4}
\urldef\tempurl%
\url{https://doi.org/10.1145/2207676.2208621}
\showDOI{\tempurl}


\bibitem[\protect\citeauthoryear{Lacerenza, Marlow, Tannenbaum, and
  Salas}{Lacerenza et~al\mbox{.}}{2018}]%
        {lacerenza2018team}
\bibfield{author}{\bibinfo{person}{Christina~N Lacerenza},
  \bibinfo{person}{Shannon~L Marlow}, \bibinfo{person}{Scott~I Tannenbaum},
  {and} \bibinfo{person}{Eduardo Salas}.} \bibinfo{year}{2018}\natexlab{}.
\newblock \showarticletitle{Team development interventions: Evidence-based
  approaches for improving teamwork.}
\newblock \bibinfo{journal}{\emph{American Psychologist}} \bibinfo{volume}{73},
  \bibinfo{number}{4} (\bibinfo{year}{2018}), \bibinfo{pages}{517}.
\newblock


\bibitem[\protect\citeauthoryear{Langfred}{Langfred}{2007}]%
        {langfred2007downside}
\bibfield{author}{\bibinfo{person}{Claus~W Langfred}.}
  \bibinfo{year}{2007}\natexlab{}.
\newblock \showarticletitle{The downside of self-management: A longitudinal
  study of the effects of conflict on trust, autonomy, and task interdependence
  in self-managing teams}.
\newblock \bibinfo{journal}{\emph{Academy of management journal}}
  \bibinfo{volume}{50}, \bibinfo{number}{4} (\bibinfo{year}{2007}),
  \bibinfo{pages}{885--900}.
\newblock


\bibitem[\protect\citeauthoryear{Larson, Wojcik, Gokhman, DeChurch, Bell, and
  Contractor}{Larson et~al\mbox{.}}{2019}]%
        {larson2019team}
\bibfield{author}{\bibinfo{person}{Lindsay Larson}, \bibinfo{person}{Harrison
  Wojcik}, \bibinfo{person}{Ilya Gokhman}, \bibinfo{person}{Leslie DeChurch},
  \bibinfo{person}{Suzanne Bell}, {and} \bibinfo{person}{Noshir Contractor}.}
  \bibinfo{year}{2019}\natexlab{}.
\newblock \showarticletitle{Team performance in space crews: Houston, we have a
  teamwork problem}.
\newblock \bibinfo{journal}{\emph{Acta Astronautica}}  \bibinfo{volume}{161}
  (\bibinfo{year}{2019}), \bibinfo{pages}{108--114}.
\newblock


\bibitem[\protect\citeauthoryear{Lewis}{Lewis}{2000}]%
        {lewis2000leaders}
\bibfield{author}{\bibinfo{person}{Kristi~M Lewis}.}
  \bibinfo{year}{2000}\natexlab{}.
\newblock \showarticletitle{When leaders display emotion: How followers respond
  to negative emotional expression of male and female leaders}.
\newblock \bibinfo{journal}{\emph{Journal of Organizational Behavior: The
  International Journal of Industrial, Occupational and Organizational
  Psychology and Behavior}} \bibinfo{volume}{21}, \bibinfo{number}{2}
  (\bibinfo{year}{2000}), \bibinfo{pages}{221--234}.
\newblock


\bibitem[\protect\citeauthoryear{Lewis}{Lewis}{2017}]%
        {lewis2017undoing}
\bibfield{author}{\bibinfo{person}{M. Lewis}.} \bibinfo{year}{2017}\natexlab{}.
\newblock \bibinfo{booktitle}{\emph{The Undoing Project: A Friendship that
  Changed the World}}.
\newblock \bibinfo{publisher}{Penguin Books, Limited}.
\newblock
\showISBNx{9780141983042}
\urldef\tempurl%
\url{https://books.google.ca/books?id=-ltNvgAACAAJ}
\showURL{%
\tempurl}


\bibitem[\protect\citeauthoryear{Mach, Dolan, and Tzafrir}{Mach
  et~al\mbox{.}}{2010}]%
        {mach2010differential}
\bibfield{author}{\bibinfo{person}{Merce Mach}, \bibinfo{person}{Simon Dolan},
  {and} \bibinfo{person}{Shay Tzafrir}.} \bibinfo{year}{2010}\natexlab{}.
\newblock \showarticletitle{The differential effect of team members' trust on
  team performance: The mediation role of team cohesion}.
\newblock \bibinfo{journal}{\emph{Journal of Occupational and Organizational
  Psychology}} \bibinfo{volume}{83}, \bibinfo{number}{3}
  (\bibinfo{year}{2010}), \bibinfo{pages}{771--794}.
\newblock


\bibitem[\protect\citeauthoryear{Maslach, Leiter, and Jackson}{Maslach
  et~al\mbox{.}}{2012}]%
        {maslach2012making}
\bibfield{author}{\bibinfo{person}{Christina Maslach},
  \bibinfo{person}{Michael~P Leiter}, {and} \bibinfo{person}{Susan~E Jackson}.}
  \bibinfo{year}{2012}\natexlab{}.
\newblock \showarticletitle{Making a significant difference with burnout
  interventions: Researcher and practitioner collaboration}.
\newblock \bibinfo{journal}{\emph{Journal of Organizational Behavior}}
  \bibinfo{volume}{33}, \bibinfo{number}{2} (\bibinfo{year}{2012}),
  \bibinfo{pages}{296--300}.
\newblock


\bibitem[\protect\citeauthoryear{Mathieu, Maynard, Rapp, and Gilson}{Mathieu
  et~al\mbox{.}}{2008}]%
        {mathieu2008team}
\bibfield{author}{\bibinfo{person}{John Mathieu}, \bibinfo{person}{M~Travis
  Maynard}, \bibinfo{person}{Tammy Rapp}, {and} \bibinfo{person}{Lucy Gilson}.}
  \bibinfo{year}{2008}\natexlab{}.
\newblock \showarticletitle{Team effectiveness 1997-2007: A review of recent
  advancements and a glimpse into the future}.
\newblock \bibinfo{journal}{\emph{Journal of management}} \bibinfo{volume}{34},
  \bibinfo{number}{3} (\bibinfo{year}{2008}), \bibinfo{pages}{410--476}.
\newblock


\bibitem[\protect\citeauthoryear{Mesmer-Magnus and DeChurch}{Mesmer-Magnus and
  DeChurch}{2009}]%
        {mesmer2009information}
\bibfield{author}{\bibinfo{person}{Jessica~R Mesmer-Magnus} {and}
  \bibinfo{person}{Leslie~A DeChurch}.} \bibinfo{year}{2009}\natexlab{}.
\newblock \showarticletitle{Information sharing and team performance: A
  meta-analysis.}
\newblock \bibinfo{journal}{\emph{Journal of applied psychology}}
  \bibinfo{volume}{94}, \bibinfo{number}{2} (\bibinfo{year}{2009}),
  \bibinfo{pages}{535}.
\newblock


\bibitem[\protect\citeauthoryear{Mueller}{Mueller}{1994}]%
        {mueller1994societal}
\bibfield{author}{\bibinfo{person}{Frank Mueller}.}
  \bibinfo{year}{1994}\natexlab{}.
\newblock \showarticletitle{Societal effect, organizational effect and
  globalization}.
\newblock \bibinfo{journal}{\emph{Organization Studies}} \bibinfo{volume}{15},
  \bibinfo{number}{3} (\bibinfo{year}{1994}), \bibinfo{pages}{407--428}.
\newblock


\bibitem[\protect\citeauthoryear{Muresan, Gonzalez‐Ibanez, Ghosh, and
  Wacholder}{Muresan et~al\mbox{.}}{2016}]%
        {muresan2016}
\bibfield{author}{\bibinfo{person}{Smaranda Muresan}, \bibinfo{person}{Roberto
  Gonzalez‐Ibanez}, \bibinfo{person}{Debanjan Ghosh}, {and}
  \bibinfo{person}{Nina Wacholder}.} \bibinfo{year}{2016}\natexlab{}.
\newblock \showarticletitle{Identification of nonliteral language in social
  media: A case study on sarcasm}.
\newblock \bibinfo{journal}{\emph{Journal of the Association for Information
  Science and Technology}} (\bibinfo{year}{2016}), \bibinfo{pages}{2725--2737}.
\newblock
\urldef\tempurl%
\url{https://doi.org/10.1002/asi.23624}
\showURL{%
\tempurl}


\bibitem[\protect\citeauthoryear{Nguyen, Shirai, and Velcin}{Nguyen
  et~al\mbox{.}}{2015}]%
        {nguyen2015sentiment}
\bibfield{author}{\bibinfo{person}{Thien~Hai Nguyen}, \bibinfo{person}{Kiyoaki
  Shirai}, {and} \bibinfo{person}{Julien Velcin}.}
  \bibinfo{year}{2015}\natexlab{}.
\newblock \showarticletitle{Sentiment analysis on social media for stock
  movement prediction}.
\newblock \bibinfo{journal}{\emph{Expert Systems with Applications}}
  \bibinfo{volume}{42}, \bibinfo{number}{24} (\bibinfo{year}{2015}),
  \bibinfo{pages}{9603--9611}.
\newblock


\bibitem[\protect\citeauthoryear{Niculae and Danescu-Niculescu-Mizil}{Niculae
  and Danescu-Niculescu-Mizil}{2016}]%
        {niculae2016conversational}
\bibfield{author}{\bibinfo{person}{Vlad Niculae} {and}
  \bibinfo{person}{Cristian Danescu-Niculescu-Mizil}.}
  \bibinfo{year}{2016}\natexlab{}.
\newblock \showarticletitle{Conversational markers of constructive
  discussions}.
\newblock \bibinfo{journal}{\emph{arXiv preprint arXiv:1604.07407}}
  (\bibinfo{year}{2016}).
\newblock


\bibitem[\protect\citeauthoryear{Niculae and
  Danescu{-}Niculescu{-}Mizil}{Niculae and Danescu{-}Niculescu{-}Mizil}{2016}]%
        {constructivedisc}
\bibfield{author}{\bibinfo{person}{Vlad Niculae} {and}
  \bibinfo{person}{Cristian Danescu{-}Niculescu{-}Mizil}.}
  \bibinfo{year}{2016}\natexlab{}.
\newblock \showarticletitle{Conversational Markers of Constructive
  Discussions}.
\newblock \bibinfo{journal}{\emph{CoRR}}  \bibinfo{volume}{abs/1604.07407}
  (\bibinfo{year}{2016}).
\newblock
\showeprint[arxiv]{1604.07407}
\urldef\tempurl%
\url{http://arxiv.org/abs/1604.07407}
\showURL{%
\tempurl}


\bibitem[\protect\citeauthoryear{Olson and Olson}{Olson and Olson}{2000}]%
        {olson2000distance}
\bibfield{author}{\bibinfo{person}{Gary~M Olson} {and}
  \bibinfo{person}{Judith~S Olson}.} \bibinfo{year}{2000}\natexlab{}.
\newblock \showarticletitle{Distance matters}.
\newblock \bibinfo{journal}{\emph{Human--computer interaction}}
  \bibinfo{volume}{15}, \bibinfo{number}{2-3} (\bibinfo{year}{2000}),
  \bibinfo{pages}{139--178}.
\newblock


\bibitem[\protect\citeauthoryear{O’Daniel and Rosenstein}{O’Daniel and
  Rosenstein}{2008}]%
        {o2008professional}
\bibfield{author}{\bibinfo{person}{Michelle O’Daniel} {and}
  \bibinfo{person}{Alan~H Rosenstein}.} \bibinfo{year}{2008}\natexlab{}.
\newblock \showarticletitle{Professional communication and team collaboration}.
\newblock In \bibinfo{booktitle}{\emph{Patient safety and quality: An
  evidence-based handbook for nurses}}. \bibinfo{publisher}{Agency for
  Healthcare Research and Quality (US)}.
\newblock


\bibitem[\protect\citeauthoryear{Pedregosa, Varoquaux, Gramfort, Michel,
  Thirion, Grisel, Blondel, Prettenhofer, Weiss, Dubourg, Vanderplas, Passos,
  Cournapeau, Brucher, Perrot, and Duchesnay}{Pedregosa et~al\mbox{.}}{2011}]%
        {scikit-learn}
\bibfield{author}{\bibinfo{person}{F. Pedregosa}, \bibinfo{person}{G.
  Varoquaux}, \bibinfo{person}{A. Gramfort}, \bibinfo{person}{V. Michel},
  \bibinfo{person}{B. Thirion}, \bibinfo{person}{O. Grisel},
  \bibinfo{person}{M. Blondel}, \bibinfo{person}{P. Prettenhofer},
  \bibinfo{person}{R. Weiss}, \bibinfo{person}{V. Dubourg}, \bibinfo{person}{J.
  Vanderplas}, \bibinfo{person}{A. Passos}, \bibinfo{person}{D. Cournapeau},
  \bibinfo{person}{M. Brucher}, \bibinfo{person}{M. Perrot}, {and}
  \bibinfo{person}{E. Duchesnay}.} \bibinfo{year}{2011}\natexlab{}.
\newblock \showarticletitle{Scikit-learn: Machine Learning in {P}ython}.
\newblock \bibinfo{journal}{\emph{Journal of Machine Learning Research}}
  \bibinfo{volume}{12} (\bibinfo{year}{2011}), \bibinfo{pages}{2825--2830}.
\newblock


\bibitem[\protect\citeauthoryear{Purdy, Nye, and Balakrishnan}{Purdy
  et~al\mbox{.}}{2000}]%
        {purdy2000}
\bibfield{author}{\bibinfo{person}{Jill~M Purdy}, \bibinfo{person}{Pete Nye},
  {and} \bibinfo{person}{PV Balakrishnan}.} \bibinfo{year}{2000}\natexlab{}.
\newblock \showarticletitle{The impact of communication media on negotiation
  outcomes}.
\newblock \bibinfo{journal}{\emph{International Journal of Conflict
  Management}} \bibinfo{volume}{11}, \bibinfo{number}{2}
  (\bibinfo{year}{2000}), \bibinfo{pages}{162--187}.
\newblock


\bibitem[\protect\citeauthoryear{Robinson, Navea, and Ickes}{Robinson
  et~al\mbox{.}}{2013}]%
        {robinson2013predicting}
\bibfield{author}{\bibinfo{person}{Rebecca~L Robinson},
  \bibinfo{person}{Reanelle Navea}, {and} \bibinfo{person}{William Ickes}.}
  \bibinfo{year}{2013}\natexlab{}.
\newblock \showarticletitle{Predicting final course performance from
  students’ written self-introductions: A LIWC analysis}.
\newblock \bibinfo{journal}{\emph{Journal of Language and Social Psychology}}
  \bibinfo{volume}{32}, \bibinfo{number}{4} (\bibinfo{year}{2013}),
  \bibinfo{pages}{469--479}.
\newblock


\bibitem[\protect\citeauthoryear{Shibani, Koh, Lai, and Shim}{Shibani
  et~al\mbox{.}}{2017}]%
        {shibani2017assessing}
\bibfield{author}{\bibinfo{person}{Antonette Shibani},
  \bibinfo{person}{Elizabeth Koh}, \bibinfo{person}{Vivian Lai}, {and}
  \bibinfo{person}{Kyong~Jin Shim}.} \bibinfo{year}{2017}\natexlab{}.
\newblock \showarticletitle{Assessing the language of chat for teamwork
  dialogue}.
\newblock \bibinfo{journal}{\emph{Journal of Educational Technology \&
  Society}} \bibinfo{volume}{20}, \bibinfo{number}{2} (\bibinfo{year}{2017}),
  \bibinfo{pages}{224--237}.
\newblock


\bibitem[\protect\citeauthoryear{Somasundaran and Wiebe}{Somasundaran and
  Wiebe}{2010}]%
        {somasundaran2010recognizing}
\bibfield{author}{\bibinfo{person}{Swapna Somasundaran} {and}
  \bibinfo{person}{Janyce Wiebe}.} \bibinfo{year}{2010}\natexlab{}.
\newblock \showarticletitle{Recognizing stances in ideological on-line
  debates}. In \bibinfo{booktitle}{\emph{Proceedings of the NAACL HLT 2010
  workshop on computational approaches to analysis and generation of emotion in
  text}}. Association for Computational Linguistics, \bibinfo{pages}{116--124}.
\newblock


\bibitem[\protect\citeauthoryear{Sprigg, Jackson, and Parker}{Sprigg
  et~al\mbox{.}}{2000}]%
        {sprigg2000production}
\bibfield{author}{\bibinfo{person}{Christine~A Sprigg}, \bibinfo{person}{Paul~R
  Jackson}, {and} \bibinfo{person}{Sharon~K Parker}.}
  \bibinfo{year}{2000}\natexlab{}.
\newblock \showarticletitle{Production teamworking: The importance of
  interdependence and autonomy for employee strain and satisfaction}.
\newblock \bibinfo{journal}{\emph{Human Relations}} \bibinfo{volume}{53},
  \bibinfo{number}{11} (\bibinfo{year}{2000}), \bibinfo{pages}{1519--1543}.
\newblock


\bibitem[\protect\citeauthoryear{Stewart}{Stewart}{2006}]%
        {stewart2006meta}
\bibfield{author}{\bibinfo{person}{Greg~L Stewart}.}
  \bibinfo{year}{2006}\natexlab{}.
\newblock \showarticletitle{A meta-analytic review of relationships between
  team design features and team performance}.
\newblock \bibinfo{journal}{\emph{Journal of management}} \bibinfo{volume}{32},
  \bibinfo{number}{1} (\bibinfo{year}{2006}), \bibinfo{pages}{29--55}.
\newblock


\bibitem[\protect\citeauthoryear{Straus}{Straus}{1994}]%
        {strausandmcgrath}
\bibfield{author}{\bibinfo{person}{J.~E.~McGrath Straus, S.}}
  \bibinfo{year}{1994}\natexlab{}.
\newblock \showarticletitle{Does the medium matter? The interaction of task
  type and technology on group performance and member reactions}.
\newblock \bibinfo{journal}{\emph{Journal of Applied Psychology}}
  \bibinfo{volume}{79} (\bibinfo{year}{1994}).
\newblock


\bibitem[\protect\citeauthoryear{Straus}{Straus}{1999}]%
        {straus1999testing}
\bibfield{author}{\bibinfo{person}{Susan~G Straus}.}
  \bibinfo{year}{1999}\natexlab{}.
\newblock \showarticletitle{Testing a typology of tasks: An empirical
  validation of McGrath’s (1984) group task circumplex}.
\newblock \bibinfo{journal}{\emph{Small Group Research}} \bibinfo{volume}{30},
  \bibinfo{number}{2} (\bibinfo{year}{1999}), \bibinfo{pages}{166--187}.
\newblock


\bibitem[\protect\citeauthoryear{Suchan and Hayzak}{Suchan and Hayzak}{2001}]%
        {suchan2001}
\bibfield{author}{\bibinfo{person}{Jim Suchan} {and} \bibinfo{person}{Greg
  Hayzak}.} \bibinfo{year}{2001}\natexlab{}.
\newblock \showarticletitle{The communication characteristics of virtual teams:
  a case study}.
\newblock \bibinfo{journal}{\emph{IEEE Transactions on Professional
  Communication}} (\bibinfo{year}{2001}), \bibinfo{pages}{174--186}.
\newblock
\urldef\tempurl%
\url{https://doi.org/10.1109/47.946463}
\showURL{%
\tempurl}


\bibitem[\protect\citeauthoryear{Tausczik and Pennebaker}{Tausczik and
  Pennebaker}{2010}]%
        {tausczik2010psychological}
\bibfield{author}{\bibinfo{person}{Yla~R Tausczik} {and}
  \bibinfo{person}{James~W Pennebaker}.} \bibinfo{year}{2010}\natexlab{}.
\newblock \showarticletitle{The psychological meaning of words: LIWC and
  computerized text analysis methods}.
\newblock \bibinfo{journal}{\emph{Journal of language and social psychology}}
  \bibinfo{volume}{29}, \bibinfo{number}{1} (\bibinfo{year}{2010}),
  \bibinfo{pages}{24--54}.
\newblock


\bibitem[\protect\citeauthoryear{Vallas}{Vallas}{2003}]%
        {vallas2003}
\bibfield{author}{\bibinfo{person}{Steven Vallas}.}
  \bibinfo{year}{2003}\natexlab{}.
\newblock \showarticletitle{Why Teamwork Fails: Obstacles to Workplace Change
  in Four Manufacturing Plants}.
\newblock \bibinfo{journal}{\emph{American Sociological Review}}
  \bibinfo{volume}{68} (\bibinfo{date}{04} \bibinfo{year}{2003}).
\newblock
\urldef\tempurl%
\url{https://doi.org/10.2307/1519767}
\showDOI{\tempurl}


\bibitem[\protect\citeauthoryear{Weidmann and Deming}{Weidmann and
  Deming}{2020}]%
        {weidmann2020team}
\bibfield{author}{\bibinfo{person}{Ben Weidmann} {and} \bibinfo{person}{David~J
  Deming}.} \bibinfo{year}{2020}\natexlab{}.
\newblock \bibinfo{booktitle}{\emph{Team Players: How Social Skills Improve
  Group Performance}}.
\newblock \bibinfo{type}{{T}echnical {R}eport}. \bibinfo{institution}{National
  Bureau of Economic Research}.
\newblock


\bibitem[\protect\citeauthoryear{Whiting, Blaising, Barreau, Fiuza, Marda,
  Valentine, and Bernstein}{Whiting et~al\mbox{.}}{2019a}]%
        {whiting2019did}
\bibfield{author}{\bibinfo{person}{Mark~E. Whiting}, \bibinfo{person}{Allie
  Blaising}, \bibinfo{person}{Chloe Barreau}, \bibinfo{person}{Laura Fiuza},
  \bibinfo{person}{Nik Marda}, \bibinfo{person}{Melissa Valentine}, {and}
  \bibinfo{person}{Michael~S. Bernstein}.} \bibinfo{year}{2019}\natexlab{a}.
\newblock \showarticletitle{Did It Have To End This Way? Understanding The
  Consistency of Team Fracture}.
\newblock \bibinfo{journal}{\emph{Proc. ACM Hum.-Comput. Interact.}}
  \bibinfo{volume}{3}, \bibinfo{number}{CSCW}, Article
  \bibinfo{articleno}{Article 209} (\bibinfo{date}{Nov.} \bibinfo{year}{2019}),
  \bibinfo{numpages}{23}~pages.
\newblock
\urldef\tempurl%
\url{https://doi.org/10.1145/3359311}
\showDOI{\tempurl}


\bibitem[\protect\citeauthoryear{Whiting, Gao, Xing, N’Godjigui, Nguyen, and
  Bernstein}{Whiting et~al\mbox{.}}{2020}]%
        {whiting2020parallel}
\bibfield{author}{\bibinfo{person}{Mark~E Whiting}, \bibinfo{person}{Irena
  Gao}, \bibinfo{person}{Michelle Xing}, \bibinfo{person}{Junior~Diarrassouba
  N’Godjigui}, \bibinfo{person}{Tonya Nguyen}, {and}
  \bibinfo{person}{Michael~S Bernstein}.} \bibinfo{year}{2020}\natexlab{}.
\newblock \showarticletitle{Parallel Worlds: Repeated Initializations of the
  Same Team To Improve Team Viability}.
\newblock \bibinfo{journal}{\emph{Proc. ACM Hum.-Comput. Interact.}}
  \bibinfo{volume}{4}, \bibinfo{number}{CSCW1}, Article
  \bibinfo{articleno}{Article 67} (\bibinfo{date}{May} \bibinfo{year}{2020}),
  \bibinfo{numpages}{23}~pages.
\newblock
\urldef\tempurl%
\url{https://doi.org/10.1145/3392877}
\showDOI{\tempurl}


\bibitem[\protect\citeauthoryear{Whiting, Hugh, and Bernstein}{Whiting
  et~al\mbox{.}}{2019b}]%
        {whiting2019fair}
\bibfield{author}{\bibinfo{person}{Mark~E Whiting}, \bibinfo{person}{Grant
  Hugh}, {and} \bibinfo{person}{Michael~S Bernstein}.}
  \bibinfo{year}{2019}\natexlab{b}.
\newblock \showarticletitle{Fair Work: Crowd Work Minimum Wage with One Line of
  Code}. In \bibinfo{booktitle}{\emph{Proceedings of the AAAI Conference on
  Human Computation and Crowdsourcing}}, Vol.~\bibinfo{volume}{7}.
  \bibinfo{pages}{197--206}.
\newblock


\bibitem[\protect\citeauthoryear{Zhang, Chang, Danescu-Niculescu-Mizil, Dixon,
  Hua, Thain, and Taraborelli}{Zhang et~al\mbox{.}}{2018}]%
        {convogoneawry}
\bibfield{author}{\bibinfo{person}{Justine Zhang}, \bibinfo{person}{Jonathan~P
  Chang}, \bibinfo{person}{Cristian Danescu-Niculescu-Mizil},
  \bibinfo{person}{Lucas Dixon}, \bibinfo{person}{Yiqing Hua},
  \bibinfo{person}{Nithum Thain}, {and} \bibinfo{person}{Dario Taraborelli}.}
  \bibinfo{year}{2018}\natexlab{}.
\newblock \showarticletitle{Conversations gone awry: Detecting early signs of
  conversational failure}.
\newblock \bibinfo{journal}{\emph{arXiv preprint arXiv:1805.05345}}
  (\bibinfo{year}{2018}).
\newblock


\end{thebibliography}

%\section{Appendix}
%\label{sec:appendix}
%\input{Appendix.tex}

\end{document}